\documentclass[cameraready]{vgtc}   

\ifpdf%
  \pdfoutput=1\relax                   
  \usepackage{graphicx}                
  \DeclareGraphicsExtensions{.pdf,.png,.jpg,.jpeg} 
\else%
  \usepackage{graphicx}                
  \DeclareGraphicsExtensions{.eps}     
\fi%

\usepackage{microtype}                 
\PassOptionsToPackage{warn}{textcomp}  
\usepackage{textcomp}                  
\usepackage{mathptmx}                  
\usepackage{times}                     
         
\usepackage{cite}                      
\usepackage{tabu}                     
\usepackage{booktabs} 
\usepackage{amsfonts}
\usepackage{amssymb}
\usepackage{amsmath}
\usepackage{tabularx}

\DeclareMathAlphabet{\mathcal}{OMS}{cmsy}{m}{n}

\usepackage[dvipsnames]{xcolor}
\usepackage{color}

\onlineid{0}

\vgtccategory{Research}

\vgtcinsertpkg

\usepackage{caption}

\graphicspath{{figs/}{./}} 

\usepackage{microtype}

\usepackage{hyperref}

\newcommand{\tool} {\mbox{{\textit{Mapper Interactive}}}}

\newcommand{\etal} {{\textit{et al.}}}

\newcommand{\wrt} {{\textit{w.r.t.}}}

\newcommand{\para}[1]{\vspace{1mm}\noindent{\textbf{#1}}}

\newcommand{\myedit}[1]{\textcolor{black}{#1}}

\newcommand {\mm}[1] {\ifmmode{#1}\else{\mbox{\(#1\)}}\fi}
\newcommand{\Rspace}        {\mm{\mathbb{R}}}
\newcommand{\Xspace}        {\mm{\mathbb{X}}}
\newcommand{\mapper}{\mm{\mathcal M}}
\newcommand{\Ucal}        {\mm{\mathcal U}}
\newcommand{\Vcal}        {\mm{\mathcal V}}

\newcommand{\Mcal} 	{\mm{\mathcal{M}}}
\newcommand{\Ncal} 	{\mm{\mathcal{N}}}
\newcommand{\denselist}{\vspace{-5pt} \itemsep -3pt\parsep=-1pt\partopsep -3pt}

\usepackage{makecell}

\usepackage{xcolor,colortbl}
\definecolor{green}{rgb}{0.1,0.1,0.1}

\definecolor{amethyst}{rgb}{0.6, 0.4, 0.8}
\definecolor{aliceblue}{rgb}{0.94, 0.97, 1.0}
\definecolor{apricot}{rgb}{0.98, 0.81, 0.69}
\definecolor{aquamarine}{rgb}{0.5, 1.0, 0.83}	
\definecolor{ashgrey}{rgb}{0.7, 0.75, 0.71}
\definecolor{asparagus}{rgb}{0.53, 0.66, 0.42}
\definecolor{babyblue}{rgb}{0.54, 0.81, 0.94}
\definecolor{babypink}{rgb}{0.96, 0.76, 0.76}
\definecolor{burlywood}{rgb}{0.87, 0.72, 0.53}
\definecolor{brightlavender}{rgb}{0.75, 0.58, 0.89}

\usepackage{lipsum,multicol}

\title{Mapper Interactive: A Scalable, Extendable, and Interactive Toolbox for the Visual Exploration of High-Dimensional Data}

\author{Youjia Zhou\thanks{e-mail: zhou325@sci.utah.edu}
\and Nithin Chalapathi\thanks{e-mail: nithin.chalapathi@utah.edu} 
\and Archit Rathore\thanks{e-mail: archit.rathore@utah.edu}
\and Yaodong Zhao\thanks{e-mail: yaodong.fs@gmail.com}
\and Bei Wang\thanks{e-mail: beiwang@sci.utah.edu}}
\affiliation{\scriptsize Scientific Computing and Imaging (SCI) Institute, University of Utah}
\abstract{
The mapper algorithm is a popular tool from topological data analysis for extracting topological summaries of high-dimensional datasets. 
In this paper, we present {\tool}, a web-based framework for the interactive analysis and visualization of high-dimensional point cloud data. 
It implements the mapper algorithm in an interactive, scalable, and easily extendable way, thus supporting practical data analysis. In particular, its command-line API can compute mapper graphs for 1 million points of 256 dimensions in about 3 minutes (4 times faster than the vanilla implementation). Its visual interface allows on-the-fly computation and manipulation of the mapper graph based on user-specified parameters and supports the addition of new analysis modules with a few lines of code. 
{\tool} makes the mapper algorithm accessible to nonspecialists and accelerates topological analytics workflows.

}

\begin{document}
\maketitle


\section{Introduction}
\label{sec:introduction}

The mapper algorithm, first introduced by Singh \etal, is a popular tool from topological data analysis (TDA) for extracting topological summaries of high-dimensional datasets in the form of simplicial complexes~\cite{SinghMemoliCarlsson2007}. 
It is rooted in the idea of ``partial clustering of the data guided by a set of functions defined on the data"~\cite{SinghMemoliCarlsson2007}. 
In many practical scenarios, the 1D skeletons of such simplicial complexes -- the mapper graphs -- serve as simple descriptions of the data and capture important information about their topological structures. 

From a theoretical perspective, researchers are actively studying the mapper algorithm and its properties (e.g.,~\cite{MunchWang2016,  CarriereOudot2018,BrownBobrowskiMunch2020}). 
From an implementational perspective, a few open-sourced tools exist that implement the mapper algorithm and support data analysis, including \textit{KeplerMapper}~\cite{VeenSaulEargle2019}, \textit{giotta-tda}~\cite{TauzinLupoTunstall2020}, \textit{Gudhi}~\cite{GUDHI2020}, and \textit{Python Mapper}~\cite{MullnerBabu2013}. 
{\tool} focuses on simultaneously addressing important aspects of the mapper algorithm involving scalability, extensibility, and interactivity in an integrated way, which differentiates it from existing implementations.

{\tool} makes the mapper algorithm accessible to nonspecialists, and those with a passing knowledge of programming concepts and TDA. 
At the same time, it gives specialists the ability to extend the system by adding analysis and visualization components in a modular fashion. 

In summary, we introduce {\tool}, a toolbox for the visual exploration of high-dimensional data. 
It comes with both a command line API for offline computation and a web-based interface for online computation of mapper graphs. 
Our contributions include:
\begin{itemize} \denselist
\item \noindent \textbf{Extendability:} We demonstrate the extendability of our toolbox via simple examples for both novice and expert users. 
\item \noindent \textbf{Interactivity:} We provide three case studies that demonstrate the strengths of {\tool} in supporting fast insight generation for well-known and new datasets. 
\item \noindent \textbf{Scalability:} 
We present a simple but effective strategy for speeding up mapper graph computations. 
Such a strategy is applicable to any mapper algorithm implementation using DBSCAN as a subroutine. 
The command line API of {\tool} computes mapper graphs for 1 million points of 256 dimensions in 3 minutes; and it is generally 3 to 6 times faster than the vanilla implementation. The GPU implementation provides an additional $2$ times acceleration for 1 million points in comparison to its CPU counterpart. 

\end{itemize}
{\tool} is open source under the MIT license and is available via Github: {\url{https://mapperinteractive.github.io/}}.

\section{Related Work}
\label{sec:related-work}
A few open-source implementations of the mapper algorithm are described in the literature.
M\"{u}llner and Babu implemented \emph{Python Mapper}~\cite{MullnerBabu2013}, which computes a mapper graph with a set of predetermined parameters. 
It contains a graphical user interface (GUI) that interfaces with the library and visualizes the resulting mapper graph. 
However, it does not provide any interactive analytic features. 

Recently, Veen and Saul created \emph{KeplerMapper}~\cite{VeenSaulEargle2019,VeenSaulEargle2019b}, a versatile and user-friendly implementation of the mapper algorithm. 
\emph{KeplerMapper} provides a few interactive capabilities in the visual encoding of a single mapper graph. For instance, users can color the nodes of a mapper graph and glean some information regarding the distribution of data points within each node.  
\emph{KeplerMapper} also includes an adapter for \emph{NetworkX}~\cite{HagbergSchultSwart2008}, where users can manually create a visualization of a mapper graph (generated by \emph{KeplerMapper}) using \emph{NetworkX}.
Similar to \emph{Python Mapper}, \emph{KeplerMapper} precomputes each mapper graph with a set of predetermined parameters; the resulting visualization is exported as a separate HTML file to be loaded in a browser.
However, its mapper implementation does not scale with a large number of points. 

The mapper algorithm is also available in the \emph{giotto-tda}  library~\cite{TauzinLupoTunstall2020}, which implements visualization capabilities as widgets within the \emph{Jupyter Notebook} environment~\cite{KluyverRagan-KelleyPerez2016}. 
Users can visualize the mapper graph in a static or an interactive mode. 
In the interactive mode, a \emph{Jupyter Notebook} widget is used to modify some of the mapper parameters; 
although the interactivity enabled {\wrt} the mapper graph object is limited. 
\emph{giotto-tda} uses \emph{Plotly}, a wrapper over \emph{D3.js} to provide some visualization capabilities.  
It focuses on creating a Pipeline object that interfaces with \emph{scikit-learn} for downstream analysis (e.g., using the mapper graph for classification or parameter tuning via a grid search).
Both \emph{giotto-tda} and \emph{Mapper Interactive} are equipped with on-the-fly computation of mapper graphs; however, the latter comes with a more  extendable implementation. 
\emph{Mapper Interactive} provides more opportunities to interact with the mapper graphs via data analysis and machine learning (ML) modules  (such as applying linear regression to a subset of nodes in a mapper graph). 
 
\emph{Gudhi}~\cite{GUDHI2020} is a TDA toolkit that contains a version of the mapper algorithm. 
It defers the visualization of the mapper graph to other tools, such as \emph{Graphviz}~\cite{EllsonGansnerKoutsofios2003}, \emph{Geomview}~\cite{PhillipsLevyMunzner1993}, and  \emph{KeplerMapper}.

\para{Comparison with the state-of-the-art.}
To further illustrate the capabilities of {\tool}, we compare its features against two of the state-of-the-art tools, namely, \emph{giotto-tda} (GT) and \emph{KeplerMapper} (KM), as shown in~\autoref{table:compare}. 
Features that are unique to {\tool} include (but are not limited to): supporting interactive parameter adjustment via a visual interface; selecting nodes from a mapper graph that form connected components or paths for comparative analysis and ML tasks; easily-extendable visual interface via low-code development; and scalable GPU  implementations for high-dimensional (100D+) point cloud data. 
In particular, {\tool} supports path selection, as certain paths in the mapper graphs have been shown to be interesting in studying the high-dimensional parameter space of plant phenomics~\cite{KalyanaramanKamruzzamanKrishnamoorthy2019}. 
Neither \emph{KeplerMapper} nor \emph{giotto-tda} supports the selection of paths or connected components of graphs for local, on-the-fly ML tasks such as dimensionality reduction or linear regression.  
Although \emph{KeplerMapper} provides certain interactivity in exploring a pre-computed mapper graph (such as selecting and displaying details for a single node), it does not allow interactive exploration of multiple parameter combinations on-the-fly. 
\emph{giotto-tda} recomputes mapper graphs based on parameters with a Python widget; however it has to regenerate HTML widgets for each dataset, and its mapper graph layout is static.
On the other hand, \emph{giotto-tda} and \emph{KeplerMapper} have their unique features. 
Both \emph{giotto-tda} and \emph{KeplerMapper} can be run as libraries inside another Python script. 
\emph{giotto-tda} outputs a Pipeline object that interfaces naturally with other \emph{scikit-learn} objects.

\begin{center}
\begin{table}[h]
\begin{tabularx}{\columnwidth}{|X|m{0.015\textwidth}|m{0.015\textwidth}|m{0.015\textwidth}|} 
\hline
Features & MI & GT & KM \\ \hline
\rowcolor{aliceblue}\multicolumn{4}{|c|}{Mapper graph computation and visualization} \\ \hline
Visualize a pre-computed mapper graph &  \cellcolor{babyblue}{Y} & \cellcolor{babypink}{N} & \cellcolor{babyblue}{Y} \\ \hline
Dynamic mapper graph layout&  \cellcolor{babyblue}{Y} & \cellcolor{babypink}{N} & \cellcolor{babyblue}{Y} \\ \hline
Compute mapper graphs via command line API & \cellcolor{babyblue}{Y} & \cellcolor{babypink}{N} & \cellcolor{babypink}{N} \\ \hline
Compute mapper graphs via a standalone GUI & \cellcolor{babyblue}{Y} & \cellcolor{babypink}{N} & \cellcolor{babypink}{N} \\ \hline
Update colormaps for continuous variables & \cellcolor{babyblue}{Y} & \cellcolor{babyblue}{Y} & \cellcolor{babyblue}{Y} \\ \hline
Enable glyphs for categorical variables & \cellcolor{babyblue}{Y} & \cellcolor{babypink}{N} & \cellcolor{babypink}{N} \\ \hline
Dynamic node size adjustment & \cellcolor{babyblue}{Y} & \cellcolor{babyblue}{Y} & \cellcolor{babypink}{N} \\ \hline
Node contraction and edge filtering in GUI &
\cellcolor{babypink}{N} & \cellcolor{babyblue}{Y} & \cellcolor{babypink}{N} \\
\hline
\rowcolor{aliceblue}\multicolumn{4}{|c|}{Node selection} \\ \hline
Select and display details for a single node & \cellcolor{babyblue}{Y} & \cellcolor{babyblue}{Y} & \cellcolor{babyblue}{Y} \\ \hline
Select and display labels for a subset of nodes & \cellcolor{babyblue}{Y} & \cellcolor{babypink}{N} & \cellcolor{babypink}{N} \\ \hline
Select nodes from a component for analysis & \cellcolor{babyblue}{Y} & \cellcolor{babypink}{\cellcolor{babypink}{N}} & \cellcolor{babypink}{N} \\ \hline 
Select nodes along a particular path for analysis & \cellcolor{babyblue}{Y} & \cellcolor{babypink}{\cellcolor{babypink}{N}} & \cellcolor{babypink}{N} \\ \hline
\rowcolor{aliceblue}\multicolumn{4}{|c|}{Data analysis and machine learning (ML)} \\ \hline
Apply ML to the entire mapper graph & 
\cellcolor{babyblue}{Y} & \cellcolor{babyblue}{\cellcolor{babyblue}{Y}} & \cellcolor{babypink}{N} 
\\ \hline
Apply ML to a selected subset of nodes & \cellcolor{babyblue}{Y} & \cellcolor{babypink}{\cellcolor{babypink}{N}} & \cellcolor{babypink}{N} \\ \hline
Compare nodes with regression & \cellcolor{babyblue}{Y} & \cellcolor{babypink}{N} & \cellcolor{babypink}{N} \\ \hline 
Compare mapper results with other ML results & \cellcolor{babyblue}{Y} & \cellcolor{babyblue}{Y} & \cellcolor{babyblue}{Y} \\ \hline
\rowcolor{aliceblue}\multicolumn{4}{|c|}{Parameter control} \\ \hline
Adjust parameters via a standalone GUI & \cellcolor{babyblue}{Y} & \cellcolor{babypink}{N} & \cellcolor{babypink}{N} \\ \hline
Adjust parameters in a \emph{Jupyter Notebook} & 
\cellcolor{babypink}{N} & \cellcolor{babyblue}{Y} & \cellcolor{babypink}{N} \\ \hline
Changing clustering approaches via \emph{scikit-learn} & \cellcolor{babyblue}{Y} & \cellcolor{babyblue}{Y} & \cellcolor{babyblue}{Y} \\ \hline
Change filter function & \cellcolor{babyblue}{Y} & \cellcolor{babyblue}{Y} & \cellcolor{babyblue}{Y} \\ \hline
\rowcolor{aliceblue}\multicolumn{4}{|c|}{Implementation and I/O} \\ \hline
Easily extensible GUI (low-code development) & \cellcolor{babyblue}{Y} & \cellcolor{babypink}{N} & \cellcolor{brightlavender}{A} \\ \hline
GPU implementation for mapper computation & \cellcolor{babyblue}{Y} & \cellcolor{babypink}{N} & \cellcolor{babypink}{N} \\ \hline
Run as a library inside Python scripts & \cellcolor{babypink}{N} & \cellcolor{babyblue}{Y} & \cellcolor{babyblue}{Y} \\ \hline
Provides \emph{scikit-learn} Pipeline object & \cellcolor{babypink}{N} & \cellcolor{babyblue}{Y} & \cellcolor{babypink}{N} \\ 
 \hline
Caching of intermediate steps with Pipeline &
\cellcolor{babypink}{N} & \cellcolor{babyblue}{Y} & \cellcolor{babypink}{N} \\
\hline
\end{tabularx}
\vspace{1mm}
\caption{Comparing features of Mapper Interactive (MI) against giotto-TDA (GT),  and Kepler Mapper (KM) respectively. Blue means ``yes'' (Y), pink means ``no" (N), purple means "almost yes" (A). The purple entry means that KM can be easily extended by users with some Python programming experience. However, MI provides more flexibilities for novice users to extend the framework by describing the new module information within a JSON file.}
\label{table:compare}
\vspace{-6mm}
\end{table}
\end{center}

\section{Background}
\label{sec:background}

We review the mapper construction introduced by Singh \etal~\cite{SinghMemoliCarlsson2007}. 
{\tool} visualizes the 1D skeleton of a mapper construction, referred to as the \emph{mapper graph}, which provides a ``skeleton-like" topological summary of a high-dimensional point cloud.  

\begin{figure}[!h]
\vspace{-2mm}
 \centering
 \includegraphics[width=.98\columnwidth]{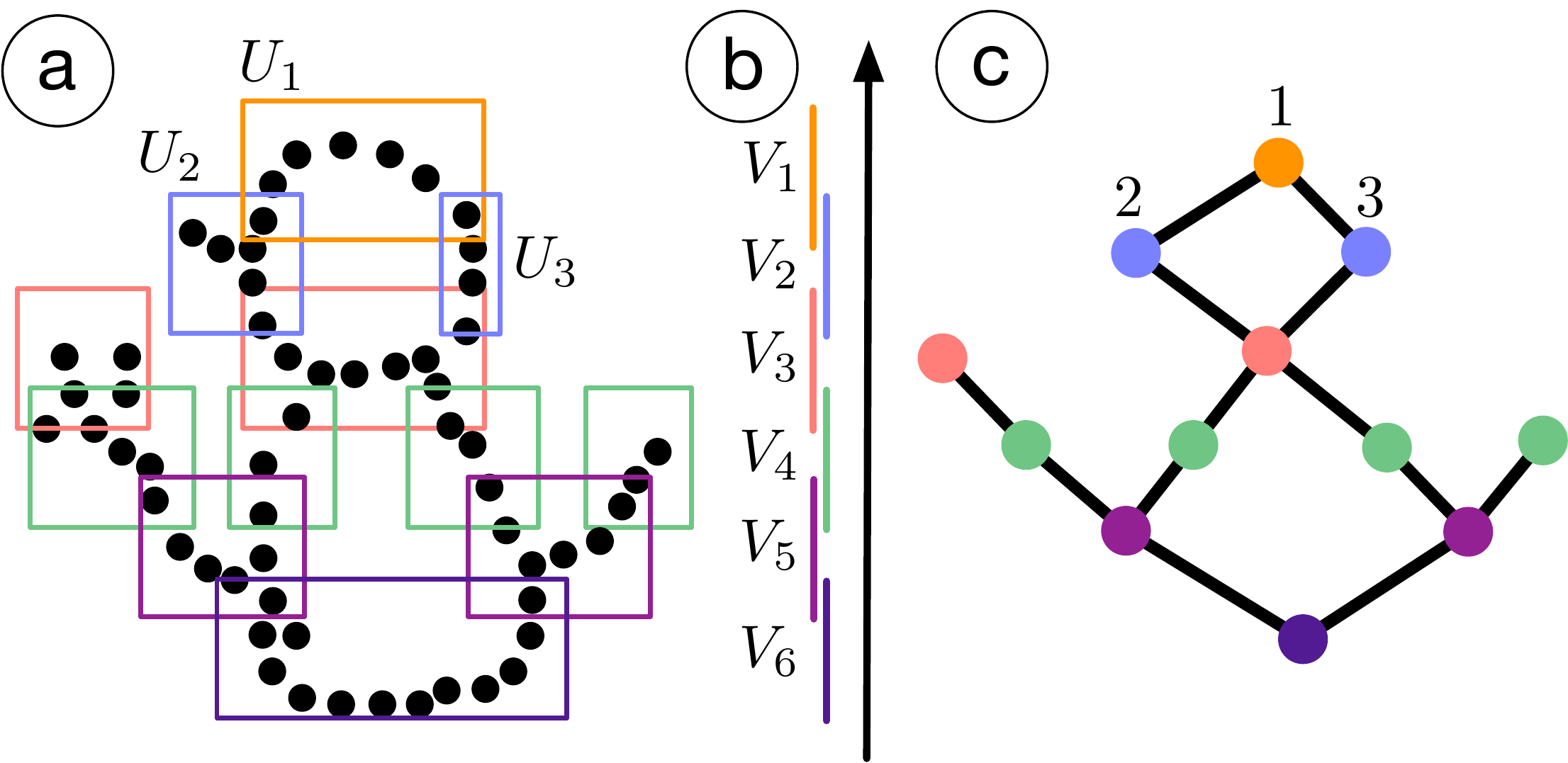}
 \vspace{-2mm}
 \caption{A mapper graph of a point cloud sampled from the silhouette of a snowman.}
 \label{fig:snowman}
 \vspace{-2mm}
\end{figure}

Given a high-dimensional point cloud $\Xspace \subset \Rspace^d$, we construct the \emph{nerve of a covering}. 
A \emph{cover} of $\Xspace$ is defined as a set of open sets in $\Rspace^d$, $\Ucal = \{U_i\}_{i \in I}$ such that $\Xspace \subset \cup_{i \in I} U_i$ ($I$ being the index set). 
The 1D nerve of $\Ucal$, denoted as $\Ncal_1(\Ucal)$, is a graph.   
Each node $i \in I$ in $\Ncal_1(\Ucal)$ represents a cover element $U_i$, and there is an edge between $i,j \in I$ if $U_i \cap U_j$ is nonempty. 
\autoref{fig:snowman}a gives an example in which $\Xspace$ is a 2D point cloud sampled from the silhouette of a snowman. 
The cover $\Ucal$ of $\Xspace$ consists a collection of rectangles on the plane. 
The 1D nerve $\Ncal_1(\Ucal)$ of $\Ucal$ is the graph in~\autoref{fig:snowman}c.  

Given a point cloud $\Xspace$, how does one obtain a cover of $\Xspace$?
In the classic mapper construction~\cite{SinghMemoliCarlsson2007}, obtaining a cover is guided by a set of scalar functions defined on $\Xspace$. 
For simplicity, we work with a single scalar function $f$ defined on $\Xspace$, $f: \Xspace \to \Rspace$. 

We start with a finite cover of a subset of the real line using intervals, that is, a cover $\Vcal = \{V_k\}$ ($1 \leq k \leq n$) of $f(\Xspace) \subset \Rspace$, such that $f(\Xspace) \subseteq \cup_{k} V_k$; see~\autoref{fig:snowman}b. 
We obtain a cover $\Ucal$ of $\Xspace$ by considering the clusters induced by points in $f^{-1}(V_k)$ for each $V_k$ as cover elements. 
The 1D nerve of $\Ucal$, denoted as $\Mcal = \Mcal(\Xspace, f):= \Ncal_1(\Ucal)$, is the \emph{mapper graph} of $(\Xspace, f)$. 

Take \autoref{fig:snowman} as an example: a point cloud $\Xspace$ is equipped with a height function, $f: \Xspace \to \Rspace$. 
Six intervals form a cover $\Vcal = \{V_1, V_2, \cdots, V_6\}$ of the image of $f$, that is, $f(\Xspace) \subset \bigcup_k V_k$. 
For each $k$ ($1 \leq k \leq 6$), $f^{-1}(V_k)$ induces some clusters that are subsets of $\Xspace$; such clusters form cover elements of $\Xspace$.
For instance, as illustrated in \autoref{fig:snowman}a, $f^{-1}(V_1)$ induces a single cluster of points that are enclosed by the orange cover element $U_1$, and $f^{-1}(V_2)$ induces two clusters of points enclosed by the blue cover elements $U_2$ and $U_3$. 
The mapper graph in \autoref{fig:snowman}c shows an edge between node $1$ and node $2$ since $U_1 \cap U_2 \neq \emptyset$. 
It captures the overall shape of the snowman.

\para{Algorithmic details in practice.}
Given a point cloud $\Xspace$, several parameters are needed to compute the mapper graph $\mapper$, including a function $f: \Xspace \to \Rspace$ (referred to as the \emph{filter function}), the number of cover elements $n$ and their percentage of overlaps $p$, the metric $d_\Xspace$ on $\Xspace$, and the clustering method. 
For example, in \autoref{fig:snowman}, $f$ is the height function, $n=6$ and $p=30\%$, $d_\Xspace$ is the Euclidean distance, and the clustering method is DBSCAN. 

In practice, the choice of the filter function $f$ is nontrivial. 
Common choices include the $L_2$-norm, variants of geodesic distances, and eccentricity~\cite{BiasottiGiorgiSpagnuolo2008, SinghMemoliCarlsson2007}. 
The mapper graph $\Mcal(\Xspace, f)$ captures the topological summary of the data $(\Xspace, f)$, that is, $\Xspace$ coupled with $f$; hence, a different choice of $f$ gives rise to a different type of summary. 
Each interval (cover element) typically has uniform size. 
Some libraries (such as \emph{giotto-tda}) offer a ``balanced" cover where the inverse image of each interval contains an equal number of points.

A common choice for the clustering method is DBSCAN~\cite{EsterKriegelSander1996}, which is a density-based clustering algorithm. 
DBSCAN has two parameters: $\epsilon$ is the neighborhood size of a given point, and $minPts$ is the minimum number of points needed to consider a collection of points as a cluster.

The filter function $f$ may be generalized to be a multivariate function, that is, $f: \Xspace \to \Rspace^m$ (for $m \geq 2$). 
In most practical scenarios $m = 2$, and the resulting mapper graph is referred to as a \emph{2D mapper graph}. The corresponding cover elements of $\Rspace^2$ become rectangles.
$\tool$ supports the computation of both 1D and 2D mapper graphs.

\section{Design and Implementation}
\label{sec:design}

We discuss three main capabilities of \tool: \emph{interactive} user interface for on-the-fly computation and exploration of mapper graphs across a range of parameters;  \emph{extendable} visualization design for novice and expert users; and a command line API that provides \emph{scalable} backend computation of mapper graphs. 

\subsection{Interactivity}

\begin{figure}[!ht]
\centering
\includegraphics[width=0.99\columnwidth]{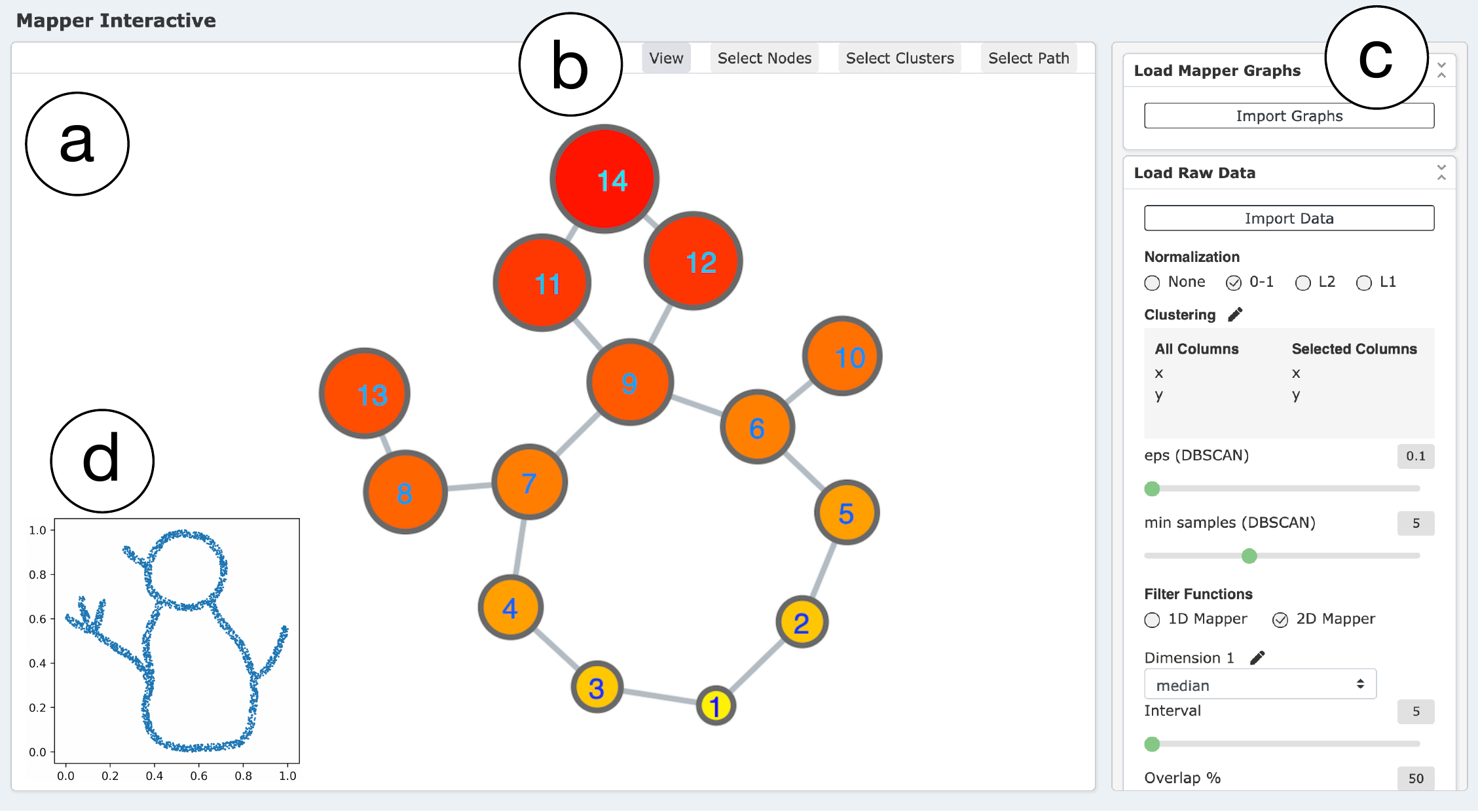}
\vspace{-2mm}
\caption{User interface of {\tool}.}
\label{fig:interface}
\vspace{-2mm}
\end{figure}

The user interface of {\tool} is shown in~\autoref{fig:interface}. It contains three main interactive panels: (a) the graph visualization panel, (b) the selection panel, and (c) the control panel.

The \textbf{graph visualization panel (a)} visualizes the resulting mapper graph using a force-directed layout, which summarizes the underlying structure of an  input point cloud dataset. 
It enables basic interactive operations such as zooming, dragging, and panning. 
In ~\autoref{fig:interface}, we see an example of a mapper graph computed from the snowman point cloud (\autoref{fig:interface}d) that appears in~\autoref{fig:snowman} of~\autoref{sec:background}.

\begin{figure}[!ht]
\centering
\includegraphics[width=0.9\columnwidth]{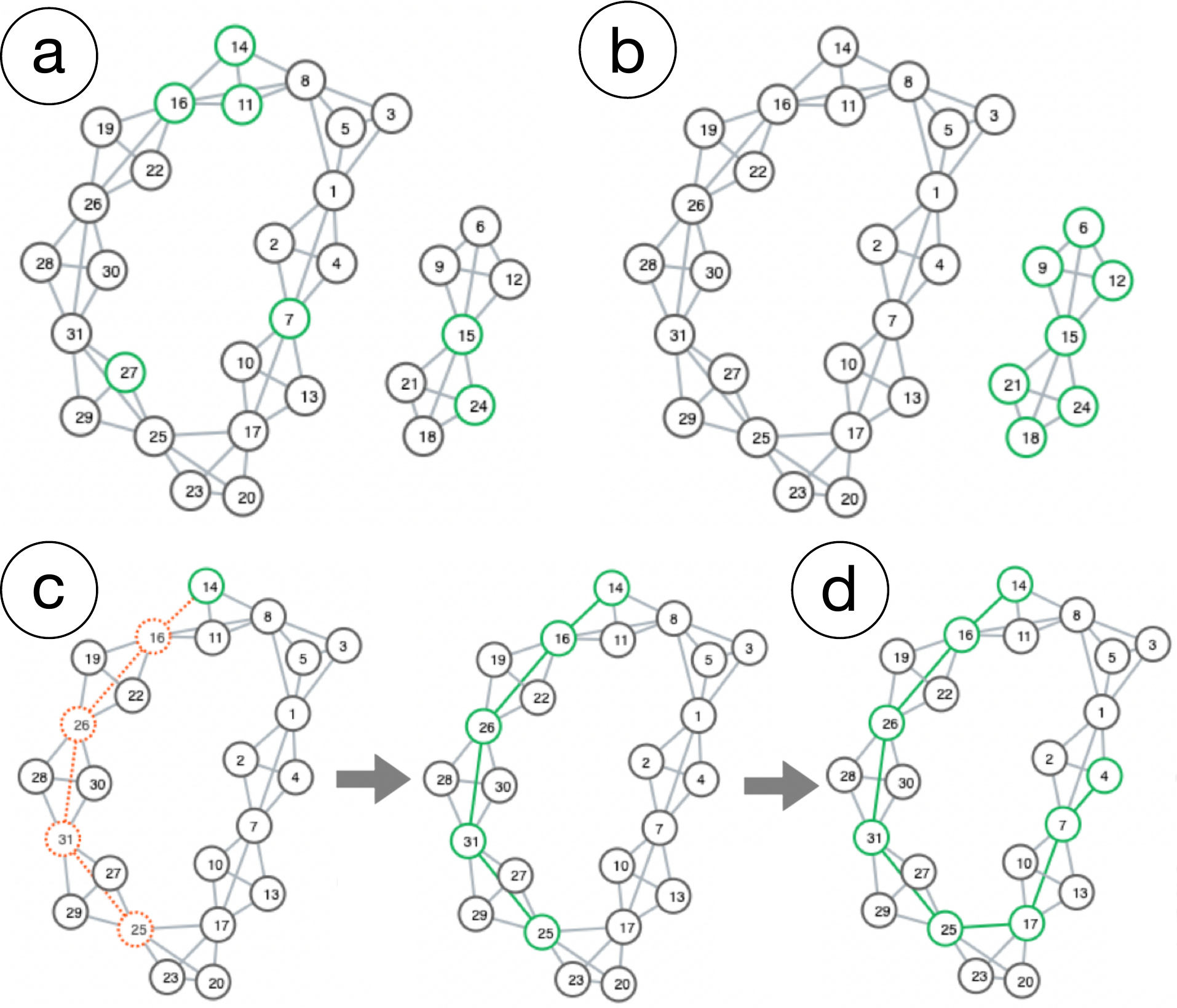}
\vspace{-2mm}
\caption{Three data selection modes for the mapper graph: (a) \textbf{select nodes}, (b) \textbf{select clusters}, and (c) \textbf{select paths} that include (d) path extensions.}
\label{fig:selection}
\vspace{-2mm}
\end{figure}

The \textbf{selection panel (b)} enables users to select a subset of mapper graph nodes (and their underlying data points) under three data selection modes. 
As illustrated in~\autoref{fig:selection}, under the \textbf{select nodes} mode (\autoref{fig:selection}a), users can select any number of the nodes in the mapper graph. 
Under the \textbf{select clusters} mode (\autoref{fig:selection}b), users can select connected components of the mapper graph. 
Under the \textbf{select paths} mode (\autoref{fig:selection}c), users can specify the start and end point of a path in the mapper graph and select a shortest path between them (if one exists). The path can also be extended by selecting another end point (\autoref{fig:selection}d).
After selection, various analysis modules can be applied to the selected data points, including linear regression and dimensionality reduction, to study the properties associated with the selected data. 

The \textbf{control panel (c)} provides parameter controls for computing mapper graphs on the fly. 
It includes data wrangling via the visual interface in addition to data wrangling provided via the command line API.  
When loading a point cloud dataset, the input data can be preprocessed through different normalization schemes such as min-max and $L_2$ normalization. 
Either 1D or 2D mapper graphs can be constructed, depending on the number of filter functions. 
A filter function can be specified based on a chosen dimension (column) of the input data, or based on derived properties from the point clouds such as $L_2$-norm, density, and eccentricity~\cite{SinghMemoliCarlsson2007}.
\myedit{For clustering, the default approach is DBSCAN. Agglomerative clustering and mean shift clustering are also available through the user interface and command line tool.}
As the parameters change, the resulting mapper graph can be computed on the fly.
In addition, the control panel interfaces with precomputed mapper graphs obtained from the command line API. 

\vspace{-2mm}

\begin{figure}[!ht]
\centering
\includegraphics[width=0.99\columnwidth]{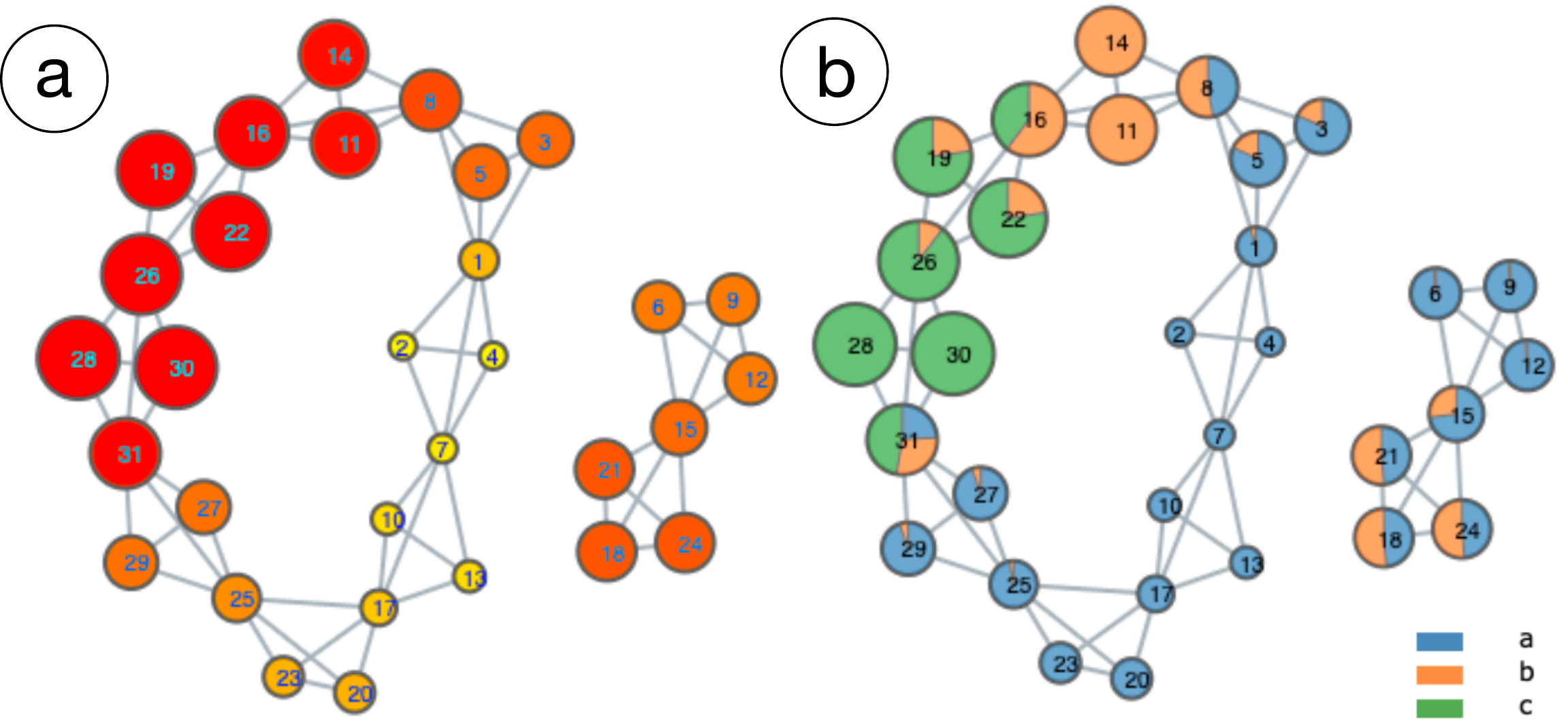}
\vspace{-2mm}
\caption{An example of the nodes colored by the average $x$ coordinates (a) and the point labels (b). The size of each node is adjusted using its average $x$ coordinate.}
\label{fig:coloring}
\vspace{-2mm}
\end{figure}

\begin{figure}[!ht]
\centering
\includegraphics[width=0.98\columnwidth]{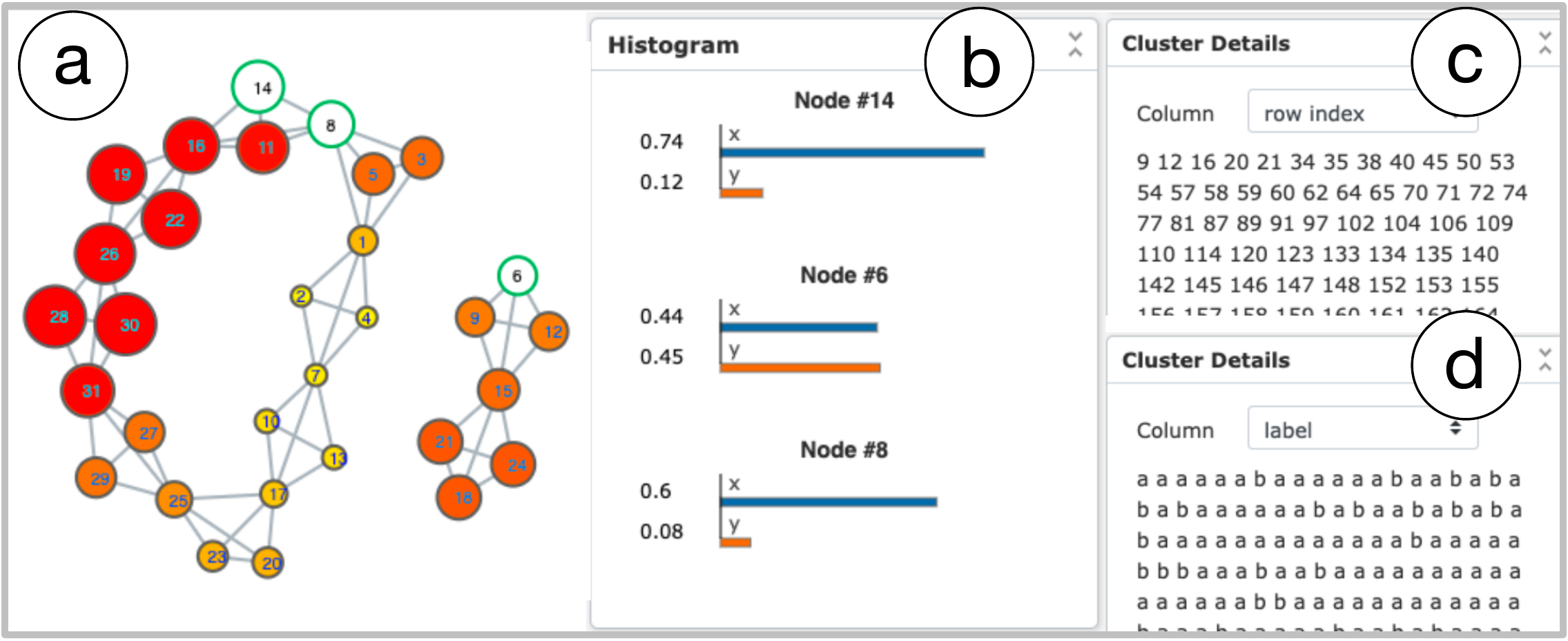}
\vspace{-2mm}
\caption{(a) In the mapper graph, nodes 6, 8, 14 are selected. (b) The bar chart of average values of $x$ and $y$ coordinates for each selected node. (c) The row indices of the union of all points contained in the selected nodes. (d) The labels of the union of all points contained in the selected nodes.}
\label{fig:selection-details}
\vspace{-4mm}
\end{figure}

The control panel also specifies parameters associated with the visual encoding of the mapper graph.  
Nodes can be colored according to a chosen dimension (variable/column) of the input data, or by the number of points contained in them. 
For discrete variables, a pie chart that reflects the composition of each node is drawn on top of each node. 
For continuous variables, a continuous colormap is applied, with user-specified color encodings and range of values. 
The size of the nodes can be adjusted using the value of a chosen variable or the number of points in the cluster (see~\autoref{fig:coloring}). 
When a subset of nodes is selected, the control panel displays the details of each node by drawing a bar chart of average values for numerical columns, and displaying the individual information of points contained in each node cluster (see~\autoref{fig:selection-details}).

\begin{figure}[!ht]
\centering
\includegraphics[width=0.9\columnwidth]{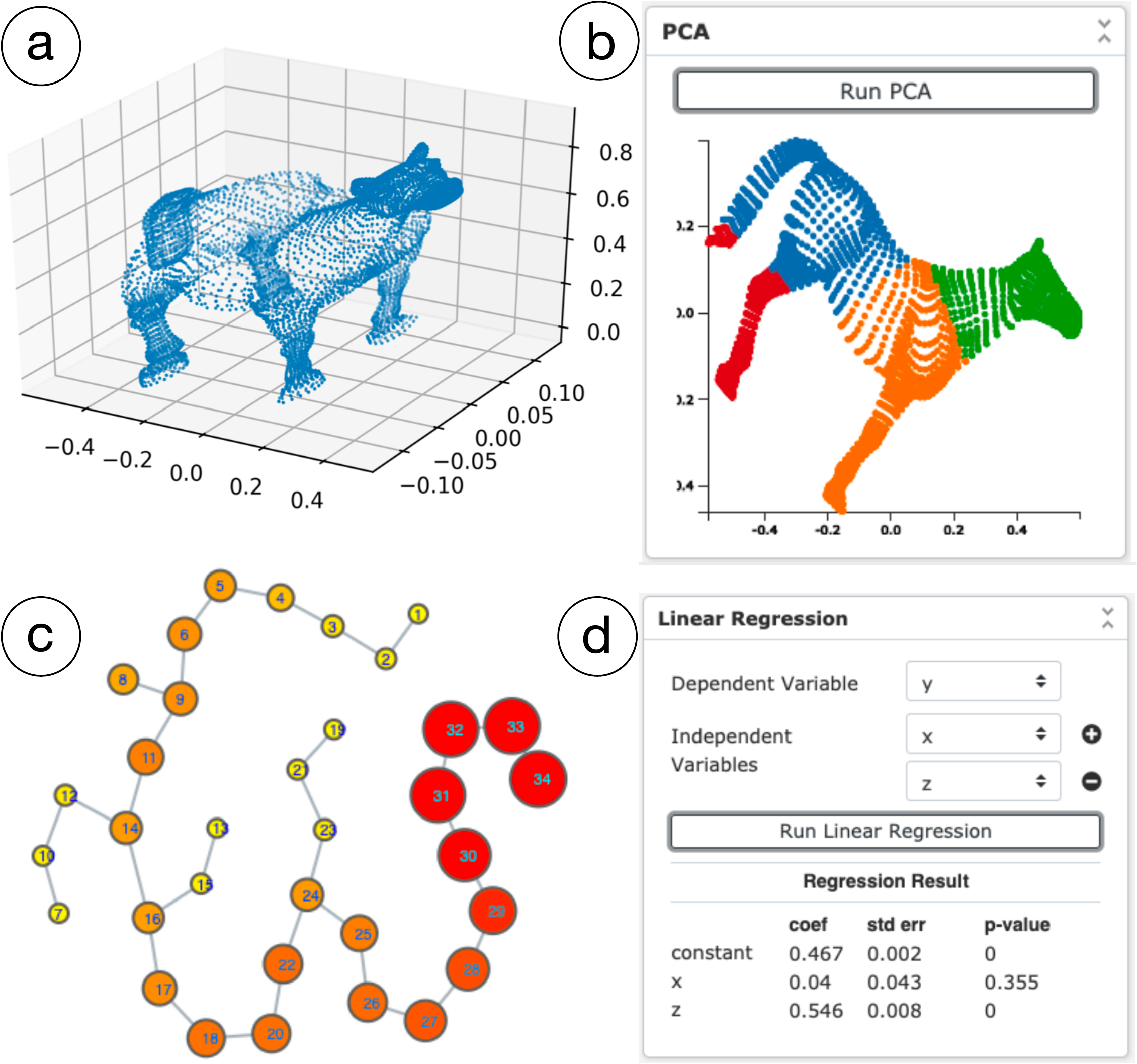}
\vspace{-2mm}
\caption{(a) A three-dimensional (3D) point cloud sampled from a model of a horse. (b) the 2D PCA result combined with $k$-means clustering where $k=4$. (c) The resulting mapper graph. (d) Linear regression result of regressing $y$ on $x$ and $z$.}
\label{fig:horse}
\vspace{-4mm}
\end{figure}

The control panel also provides data analysis and machine learning modules for users to better understand the results of the mapper algorithm, which is currently not possible with other existing tools.
Machine learning techniques, including linear regression and principal component analysis (PCA), can be applied to analyze a selected subset of nodes. If no nodes is selected, the entire dataset will be taken as input.
Take~\autoref{fig:horse} for example: (a) shows a 3D point cloud sampled from the model of a horse; (b) is the 2D PCA result with k-means clustering applied to the projected data, where colors represent different clusters; (c) is the mapper graph of (a) generated by {\tool} with nodes $1$, $7$, $13$, $19$ at the four feet of the horse, node $8$ at its tail, and node $34$ at its head; and (d) shows the results of applying linear regression to the point cloud, where $x$ and $z$ are the independent variables, and $y$ is the dependent variable. 

\subsection{Extendability}

{\tool} allows users to easily extend the framework by adding new data analysis and visualization modules to the control panel, primarily via interfacing with Python's \emph{scikit-learn} package. 
Such extendability brings flexibility for users to apply machine learning techniques to nodes (clusters) of interest that arise from the mapper graph. 
It also enables users to explore the properties associated with these nodes. 

We provide two modes for different user groups to extend the framework: the novice user mode and the expert user mode.

\para{Novice user mode.} 
For users with limited programming experience, we provide an easy way for them to add new modules. 
All they need to do is to describe the new module information within the \textsf{new\_modules.json} file, and the system will detect and generate all the new modules inside that json file automatically. 
Currently, {\tool} allows the addition of supervised and unsupervised learning algorithms that are available via \emph{scikit-learn}. 
For each new module, users need to specify the function name, function parameters, and whether it is a supervised or unsupervised model for the Python backend to fit the model correctly, along with a list of visual component types for the JavaScript frontend to visualize the result. 
For a supervised learning module, users need to provide additional information about the independent and dependent variables. 
For visualization purpose, we provide commonly used visual components, such as scatter plots, line graphs, and tables, to be integrated with the result of a new module. 

\vspace{-4mm}
\begin{figure}[!ht]
\centering
\includegraphics[width=0.99\columnwidth]{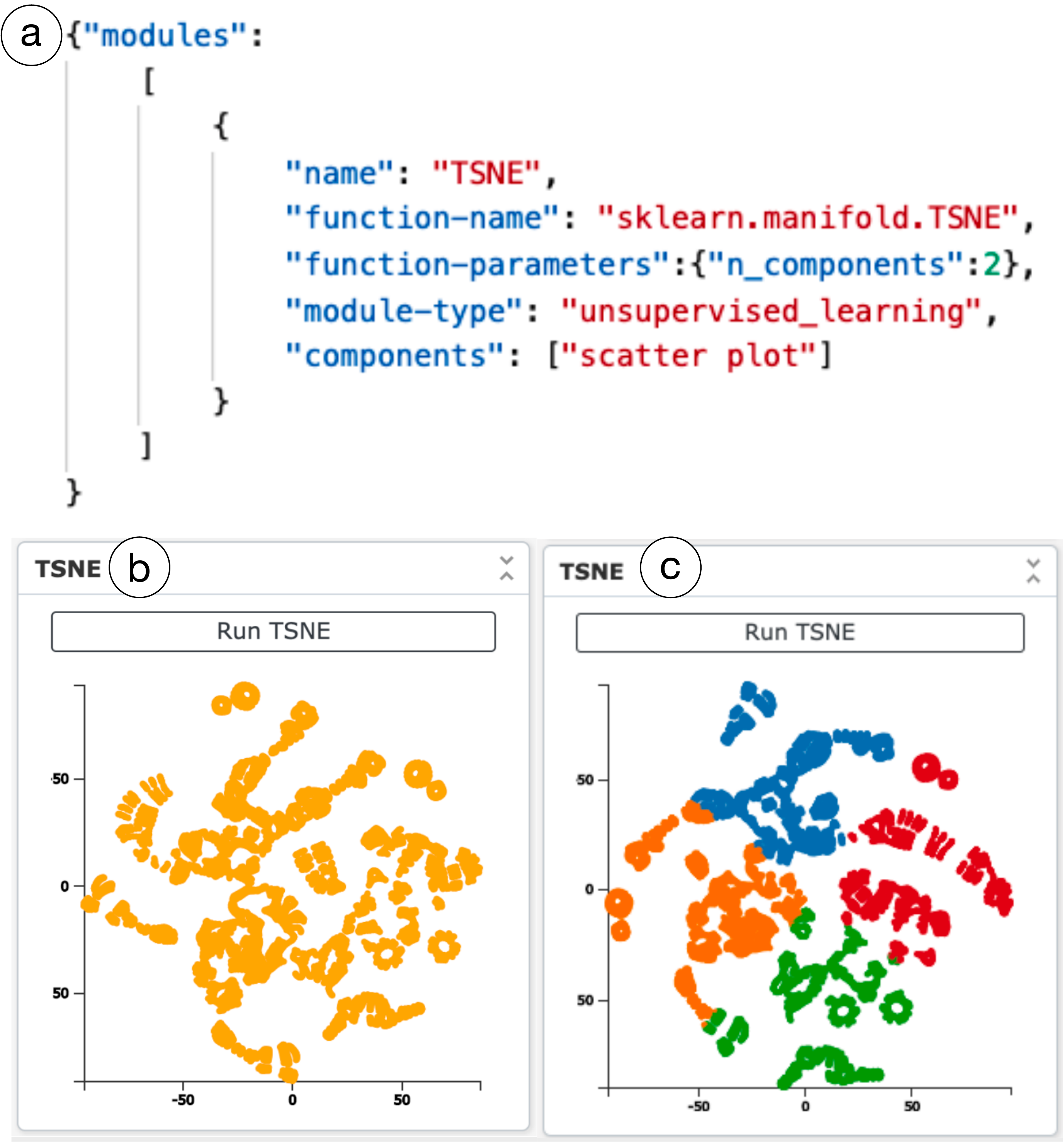}
\vspace{-2mm}
\caption{An example of adding t-SNE module to the control panel of {\tool}.}
\label{fig:extendability}
\vspace{-4mm}
\end{figure}

\para{Expert user mode.} 
For expert users with programming experience, we provide a template function \textsf{call\_module\_function} in Python within {\tool}.  
It supports customizable and multistep analysis pipelines.  
We also provide a template class in JavaScript for creating new visual components using \emph{D3.js}. 
With a few lines of code, users can add a new drawing method within the template class to modify a visual encoding. 
The styles of visual components are changed via a CSS file. 

We give an example in~\autoref{fig:extendability} for adding a t-SNE module to the {\tool} interface. t-SNE is a nonlinear dimensionality reduction technique that is quite popular in practice. 
In (a), under the novice user mode, the information of a new module that performs t-SNE-based dimensionality reduction is added to the \textsf{new\_modules.json} file. 
With less than six lines of code, the t-SNE module is now part of the {\tool} interface where a 3D point cloud from a horse is visualized in 2D (b).
In (c), under the expert user mode, by adding a few lines of code to the Python script \textsf{call\_module\_function}, the t-SNE result can be further enhanced using k-means clustering (where $k=4$). 

\subsection{\myedit{Scalability}}
{\tool} is equipped with scalable backend computation of mapper graphs. 
In particular, it comes with a command line API for data wrangling and the computation of mapper graphs. 
The single processor CPU implementation of the mapper algorithm in {\tool} is $3$ to $6$ times faster than its vanilla implementation. 
We also provide a GPU implementation that provides an additional $2$-fold speedup for 1 million points in 128 dimension.  

\para{Key implementational idea.} 
We present a simple but effective strategy for speeding up \emph{any} mapper algorithm framework that uses DBSCAN as a subroutine.  

The backend mapper implementation of {\tool} is built upon \emph{KeplerMapper}. 
\emph{KeplerMapper} is a user-friendly implementation of the mapper algorithm that provides some interactive capabilities. 
However, its default mapper graph computation (considered as the vanilla implementation) does not scale well with the size of the point cloud. 
The computational bottleneck happens during the DBSCAN clustering stage in which the algorithm queries all pairwise distances. 
The first idea is parallelizing individual clustering instances, that is, computing the clusters for the inverse map of each interval. 
This strategy provides some amount of speed-up, which is employed by both {\tool} and \emph{giotto-tda}.  

In \tool, we push the parallelization even further. We modify the algorithm by precomputing the distance matrix of points within each interval using \emph{scikit-learn}'s highly optimized \texttt{pairwise\_distance} function. 
This function converts the distance computations in the clustering algorithm to a lookup in the precomputed matrix, achieving significant speed-up at the cost of higher memory usage entailed by storing the precomputed distances. 
For {\tool}, such a strategy is shown to be 6 times faster than the vanilla implementation for 300K points for \emph{ImageNet} dataset. 

In fact, the strategy employed by {\tool} is applicable to \emph{any} mapper framework employing DBSCAN as a clustering subroutine. 
By enforcing pre-computation of distance matrices as a user-specified parameter in DBSCAN, our strategy also helps to speed up mapper graph computation for both \emph{giotto-tda} and \emph{KeplerMapper}, when the point clouds are of significantly high dimension (100D+). 
It is also important to point out that precomputing distances ceases to be  effective when the pullback cover sets become too large. 
It also does not lead to significant speed-up when the dimensionality is not sufficiently large. 

In the experiments below, we perform runtime analysis for {\tool}, \emph{KeplerMapper} (version 1.2.0) and \emph{giotto-tda} (version 0.3.1), where GT and KM represent their default configurations, respectively. 
To demonstrate that our strategy will speed up any DBSCAN-based mapper framework, we give performance numbers for GT* and KM*, which represent the improved configurations using the strategy of {\tool} in precomputing distances, respectively\footnote{Via detailed discussions with a giotto-tda developer Umberto Lupo.}. 

\para{Datasets.}
The \textit{ImageNet} and \textit{Cifar} datasets are created by passing input images to \textit{InceptionV1} and \textit{ResNet-18} neural networks respectively and collecting the activation vectors at an intermediate layer.
The \textit{ImageNet} dataset has 512 dimensions and 300K points, while the \textit{Cifar} dataset has 256 dimensions and 3 million points. 
For the \textit{Random} vector datasets, we sample 10 million points of 128 dimensions; each dimension is drawn from a uniform distribution over $[0, 1]$. 
For our experiments, we subsample each of the datasets at various order of magnitudes to demonstrate run time performance of each method at different input sizes.

\begin{figure}[!ht]
    \centering
    \includegraphics[width=0.8\linewidth]{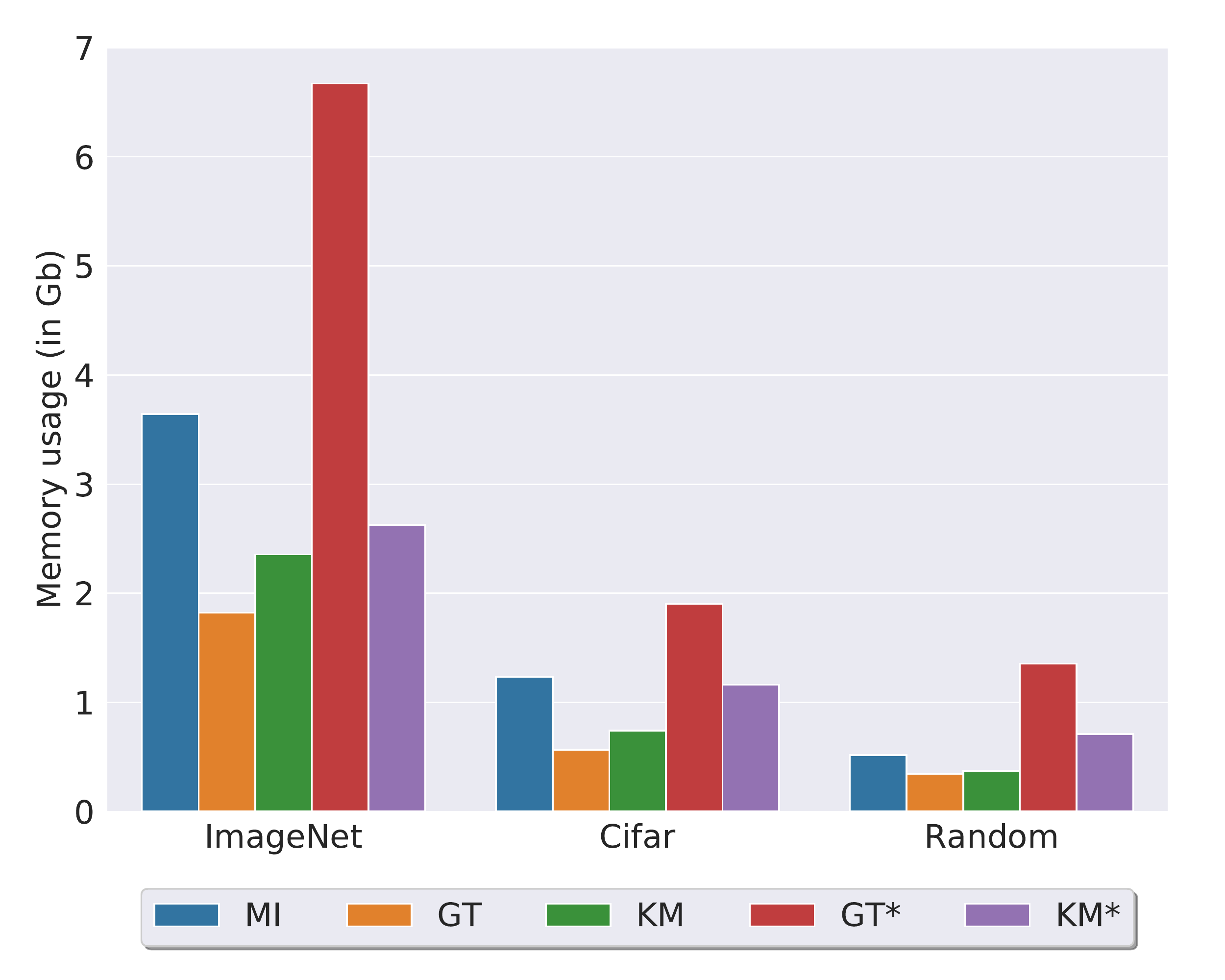}
    \vspace{-4mm}
    \caption{Peak memory usage (in Gigabytes) on three datasets, each with 100K points. The number of intervals is set to 100. The \emph{ImageNet}, \emph{CIFAR}, and \emph{Random} datasets contain $512$, $256$, and $128$ dimensional point clouds respectively.}
    \label{fig:memory-comparison}
    \vspace{-4mm}
\end{figure}

\para{Runtime analysis with the command line API.} 
We first show comparisons of peak memory usage among {\tool} (MI), GT, KM, GT*, and KM* in~\autoref{fig:memory-comparison}, the numerical values can be found in~\autoref{table:memory-comparison} of the supplement.   
Roughly speaking, for all three datasets, {\tool} has up to $1.7\times$ increase in peak memory usage compares to its vanilla implementation. Using our strategy, GT* has up to $4.0\times$ memory increase over GT; and KM* has up to $1.9\times$ memory increase over KM.   

The runtime comparison is shown in~\autoref{fig:runtime-all}(a-c) and  \autoref{table:runtime-all} in the supplement. 
By employing space-time tradeoff of MI via the pre-computation of distance matrices, both GT and KM can be improved to achieve comparable performance with MI. 
For the \emph{ImageNet} dataset of 300K points, MI achieves an approximately $6\times$ speedup against its vanilla implementation (KM); with our strategy, GT* gets $9 \times$ speedup against GT, and KM* achieves $7\times$ speedup {\wrt} to KM. 
For the \emph{Cifar} dataset of 3 million points, MI achieves a $4\times$ speed-up against its vanilla implementation; while GT* gets a $3\times$ speed-up against $GT$, and KM* obtains a $5\times$  speed-up against KM.
For the \emph{Random} dataset of 10 million points, our command line API computes a mapper graph in 29 minutes, obtaining a $3\times$ speed-up against its vanilla implementation; while KM* achieves $2\times$ speed up vs KM, and GT* and GT run out of memory.

\begin{figure*}[!ht]
	\centering
	\includegraphics[width=\linewidth]{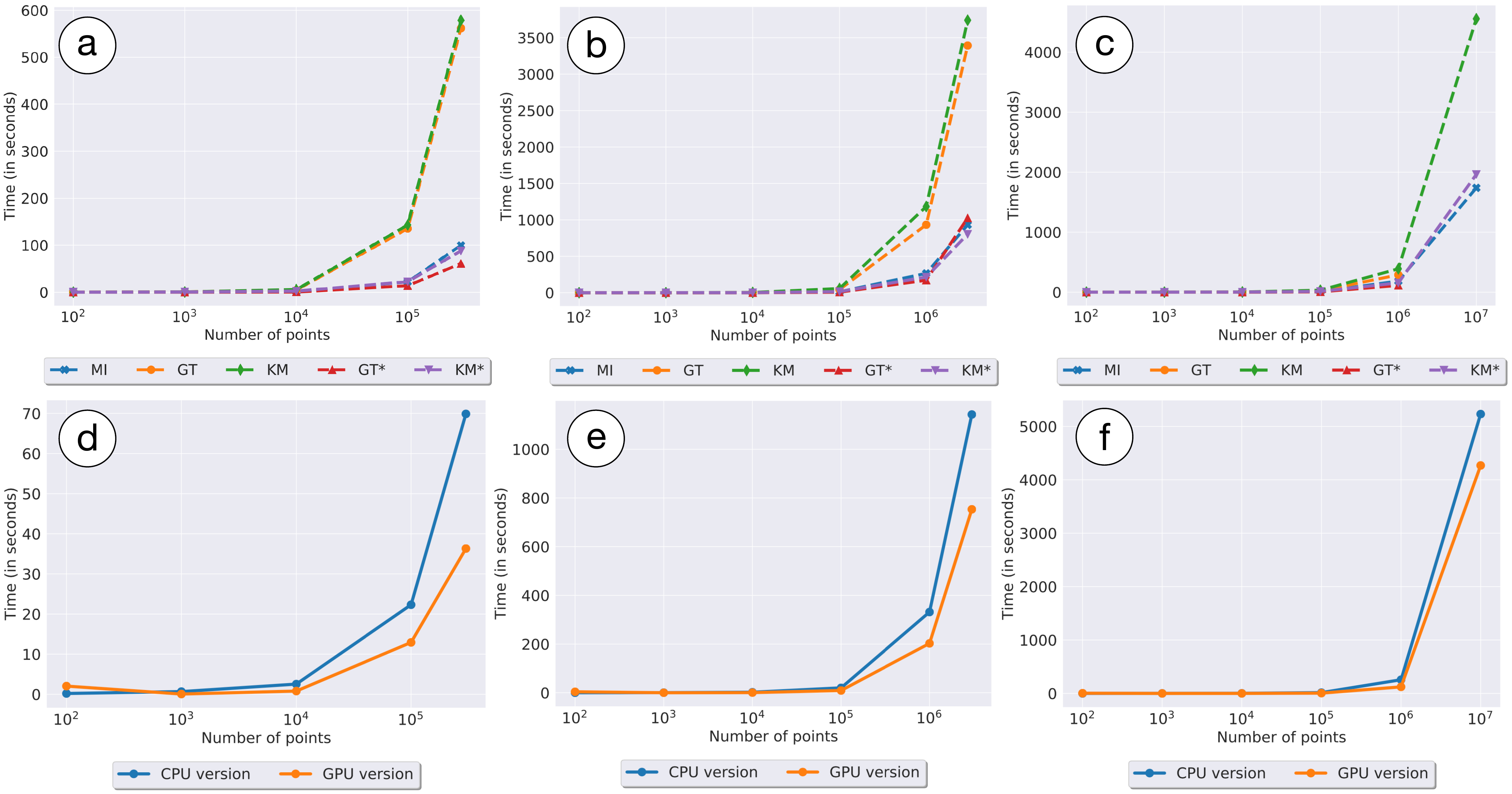}
	\vspace{-7mm}
	\caption{Top row (a-c): CPU runtime (in seconds) on the \emph{ImageNet}, \emph{Cifar}, and \emph{Random} datasets, respectively. 
      Bottom row (e-f): GPU runtime (in seconds) in comparison with CPU runtime for our implementation. Labels are shown on the x-axis using a log scale.}
	\label{fig:runtime-all}
	\vspace{-4mm}
\end{figure*}

We perform the above experiments on an Intel Xeon 2.4GHz CPU with 16 cores and 32 GB RAM. 
For parallel computations, we restrict the methods to 8 cores to minimize effects from OS processes. 
We set the \texttt{n\_jobs} parameters for the clustering algorithm in \emph{KeplerMapper} to $8$ and the same parameter for the \emph{giotto-tda} implementation to $8$ as well, to compare against the parallelized versions of our implementations.

\para{Runtime analysis with the visual interface.}
Additional I/O and memory overhead is associated with computing the mapper graph via the visual interface in comparison with the command line API. 

To test the scalability of the visual interface, we use a macOS system on an Intel 2.3 GHz Core i5 CPU and 8 GB RAM. For a dataset with 100K points in 512 dimensions, it takes an average $3$ minutes to compute and render the mapper graph in the browser. 
When the number of points increases to 200K, the computation and renders takes an average of 1 hour. 
On the other hand, if we generate the mapper graphs with 1 million points using the command line API, the interface can easily load the resulting mapper graphs in under $1$ minute. 
By interfacing with the command line API, we are able to explore larger point cloud data via the visual interface with precomputed mapper graphs.

\para{GPU accelerated distance computation.} 
Finally, in order to further speed up our parallel mapper algorithm, we introduce a GPU-based distance computation. We use \emph{PyTorch} to accelerate the distance computation, moving away from \emph{scikit-learn}. Our results are based on a computer with a 32-core Intel Xeon CPU (1.8 GHz), 132 Gb of RAM, and a Nvidia Titan V GPU with CUDA 10.1. 
We notice a roughly $2\times$ speed-up in comparison with our CPU implementation for our larger point clouds, see~\autoref{fig:runtime-all}(c-f) (numerical values are shown in~\autoref{table:GPU-all} of the supplement).
In particular, for 1 millions points, our GPU implementation achieves a  $1.6\times$, and $2.1\times$ speed-up for the \emph{Cifar} and the \emph{Random} dataset, respectively. 

In summary, the backend GPU implementation of mapper graph computation achieves (on average) between $6\times$ to $12\times$ speed-up in comparison with the CPU-based vanilla implementation. 
However, a communication overhead is incurred when large arrays of data are moved from CPU to GPU and back.
As a result, experiments with a higher number of intervals (500+) do not  provide as large of a speed-up.

\subsection{Implementation}

The visual interface of {\tool} is implemented using HTML/CSS/JavaScript stack with \emph{D3.js} and \emph{JQuery} JavaScript libraries. 
It interfaces with a Python backend via a \emph{Flask} server. 
The mapper graph computation is modified from the \emph{KeplerMapper} implementation.
Python libraries, including \emph{scikit-learn}, \emph{statsmodels}, and \emph{scipy}, are used for its machine learning modules.

We also provide a Python command line API which is designed for data wrangling and offline mapper graph computations for large datasets. The wrangling process handles missing values, identifies numerical and categorical columns, and removes non-numerical elements from the numerical columns. 
The wrangled data may be imported to the visual interface for interactive exploration. 
To compute mapper graphs via the API, users can specify the range of parameters for the mapper algorithm, including the number of intervals, the amount of overlap, the number of threads to use when computing pairwise distances, parameters for DBSCAN, etc. 
Users can also specify GPU acceleration for computation via the command line API. 
The resulting mapper graphs are put into a single folder to be interfaced with the visual interface for interactive exploration. 

\section{Use Cases}
\label{sec:use-cases}

We demonstrate the utility of {\tool} via three use cases on well-known and new datasets from image classifiers, breast cancer, and COVID-19. 
By applying {\tool} to these datasets, we showcase the usability, interactivity and extensibility of the tool. 
While a subset of the findings in these use cases may be obtained using existing frameworks, the main advantage of {\tool} is that it provides the largest set of features unavailable with existing frameworks (cf.~\autoref{table:compare}), and it makes mapper algorithm accessible to non-specialists (with little background in Python and TDA) via a low-code development environment.
In addition, with zero or a few lines of code, {\tool} enables a quick way to check if TDA is a viable tool for a given application.

Through these use cases, we demonstrate that {\tool} provides fast and easy ways to prototype and experiment with user-specified datasets, thus helping accelerate TDA workflows for fast insight generation.

\subsection{Discovering the Divergence of COVID-19 Trends}
Our first use case is to analyze and compare COVID-19 trends in the United States. 
The key point is that {\tool} enables fast insight generation on a new dataset. 

\vspace{-2mm}
\begin{figure}[!ht]
\centering
\includegraphics[width=0.98\columnwidth]{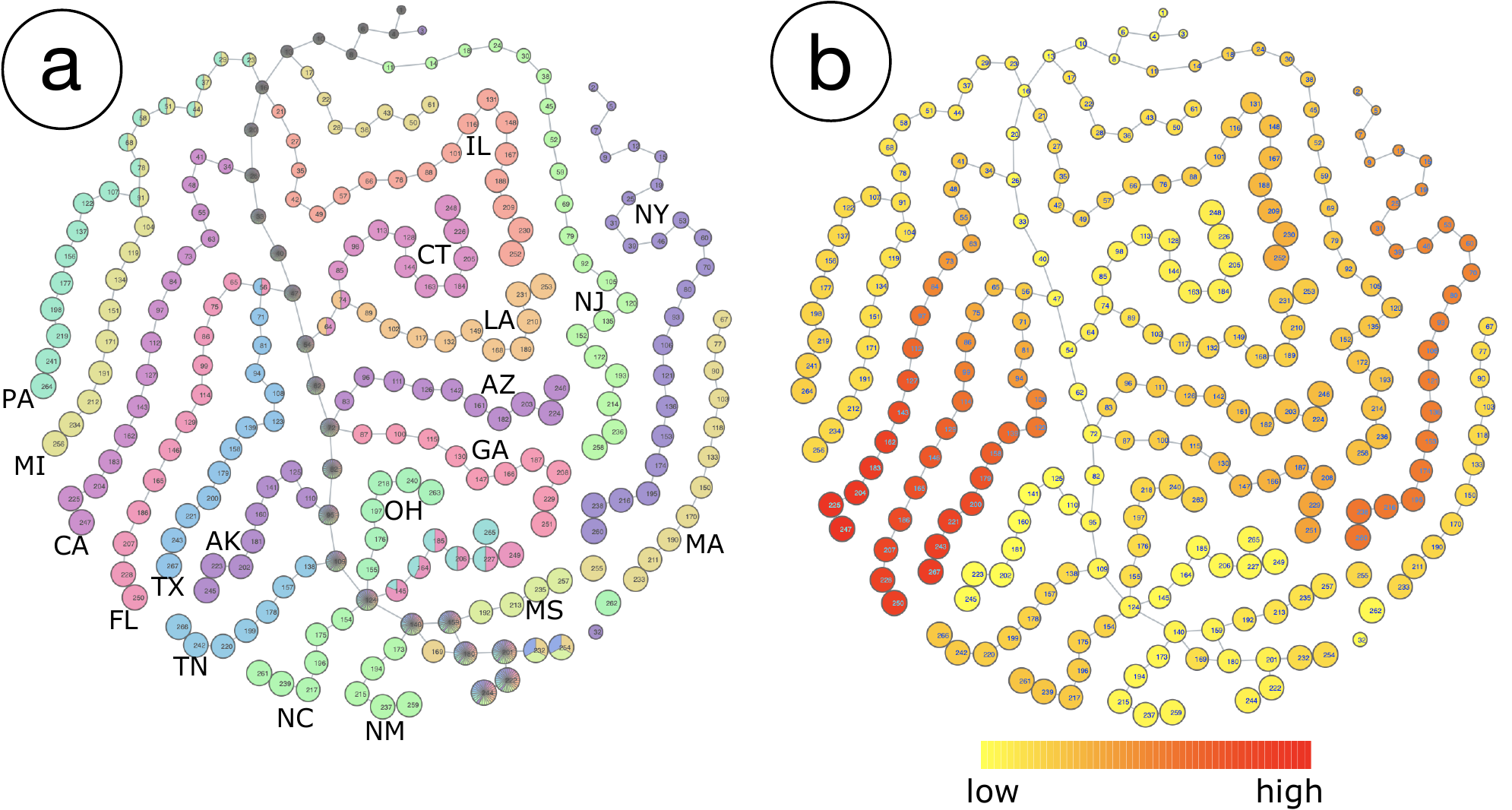}
\vspace{-2mm}
\caption{The mapper graph of the full COVID-19 dataset. The graph nodes are colored by the composition of the states (a) and the number of confirmed cases (b), respectively. For DBSCAN, we chose  $\epsilon = 0.1$, $minPts = 5$. For mapper graph, we set $n=35$ and $p=50\%$. Each dimension's values are normalized to be between 0 and 1. The size of nodes indicates the average number of recorded days.}
\label{fig:covid-all}
\vspace{-3mm}
\end{figure}

The dataset contains the daily records of COVID-19 cases in all 50 states from April 12, 2020 to September 18, 2020\footnote{\url{https://github.com/CSSEGISandData/COVID-19/}}. 
It contains 9240 data points (rows), each of which corresponds to a daily record for a given state. 
For each state, it contains 7 statistical measures (columns): number of confirmed cases, death cases, active cases, people tested, as well as the testing rate, mortality rate, and incidence rate (i.e.,~the number of cases per 100K persons).  

We first compute an initial mapper graph using all data points. 
We include all 7 dimensions (columns) to compute the pairwise distance matrix, and use the number of recorded days (since April 12, 2020) as the filter function. 
The number of recorded days indicates how many days have passed from the record starting date (April 12, 2020) to the date associated with  each row of data.

The result is shown in~\autoref{fig:covid-all}. 
Certain states are shown to be separated from other states and form their own connected components, such as New York (NY) and Massachusetts (MA), which implies that their statistics (and thus ``epidemic trends") may be quite different from others. 
To further investigate why and how these states are separated from one another, we select nine states (AZ, CA, FL, GA, IL, NC, NJ, NY, TX) with the largest number of confirmed cases and compute a second mapper graph.  

\vspace{-2mm}
\begin{figure}[!ht]
\centering
\includegraphics[width=0.8\columnwidth]{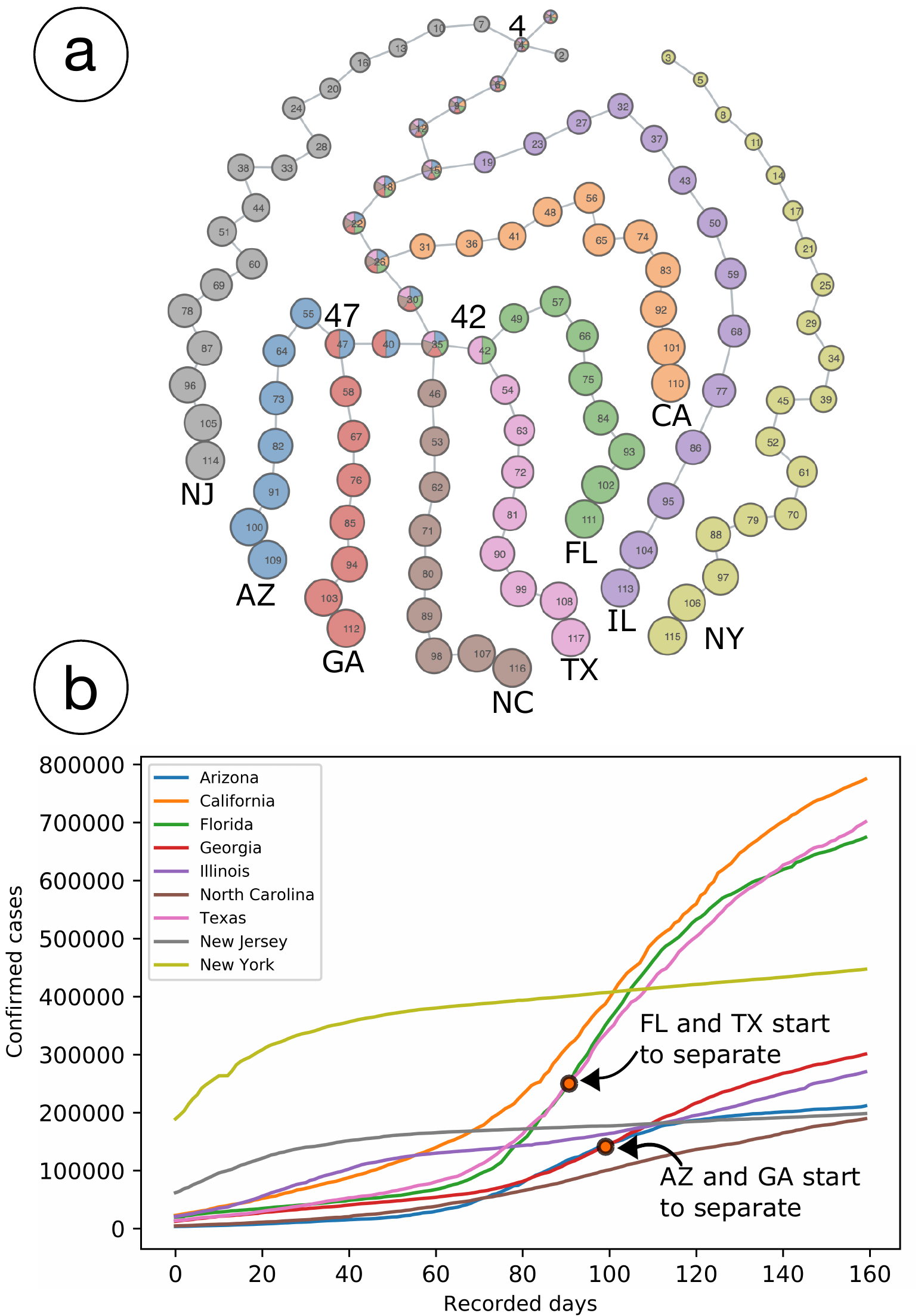}
\vspace{-3mm}
\caption{(a) The mapper graph for the selected states. (b) The line graph of the daily confirmed cases; the x-axis represents the number of recorded days, and the y-axis represents the confirmed cases. For DBSCAN, $\epsilon=$ 0.15, $minPts=5$. For the mapper graph: $n=20$, $p= 50\%$. Each dimension is normalized by a min-max scale.}
\label{fig:covid-sub}
\vspace{-3mm}
\end{figure}

As shown in~\autoref{fig:covid-sub}a, the states become separated from each other after certain branching points. 
The size of each node is encoded by the average number of recorded days. 
By comparing the line graph of the confirmed cases (\autoref{fig:covid-sub}b), we can see that the order by which each state is separated is related to how different its curve is from that of other states. 

For example, as shown in the line graph in~\autoref{fig:covid-sub}b, the curve of New York (NY) deviated the most from other states, so in the mapper graph (\autoref{fig:covid-sub}a) it is not connected with any other state, thus forming its own connected component.  
New Jersey (NJ) shows the second highest deviation besides New York in the line graph, so it splits from the main branch at node 4 in the mapper graph. 

Arizona (AZ) and Georgia (GA), as well as Florida (FL) and Texas (TX), are two pairs with similar trends in~\autoref{fig:covid-sub}b, so their nodes are the last ones to be separated from the main branch in~\autoref{fig:covid-sub}a. In particular, the average number of days at node 47 is 100, which reflects exactly where Arizona and Georgia start to separate in the line graph. 
The average number of days at node 42 is 92, which reflects when Florida and Texas start to separate in the line graph. 

Therefore, through the resulting mapper graph, we are able to distinguish states with different epidemic trends and to determine how different their trends are. We can also discover when their trends start to diverge by looking at the nodes at the branching points.

\subsection{Visualizing Class Separation via Neuron Activations}
Our second use case is to visualize neuron activations collected at the last layer of an image classifier to study the degree to which classes are separated during training. 
The key point is that {\tool} helps to highlight class separation with a categorical dataset, and it can be extended to perform in-depth analysis of the data.   

The \emph{Cifar} dataset is created by passing input images from CIFAR-10~\cite{KrizhevskyHintonothers2009} to \emph{ResNet-18} neural network, and collecting activation vectors (that is, combinations of neuron firings) from the last layer (referred to as ``4.1.bn2") of the network. 
We then treat these activation vectors as a dataset containing high-dimensional points and apply {\tool} to it.  
We use $50K$ images from $10$ image classes, namely ship, truck, automobile, horse, deer, bird, dog, cat, fog, and airplane. 
Each image corresponds to an activation vector with 512 dimensions. 

\begin{figure}[!ht]
\centering
\includegraphics[width=0.95\columnwidth]{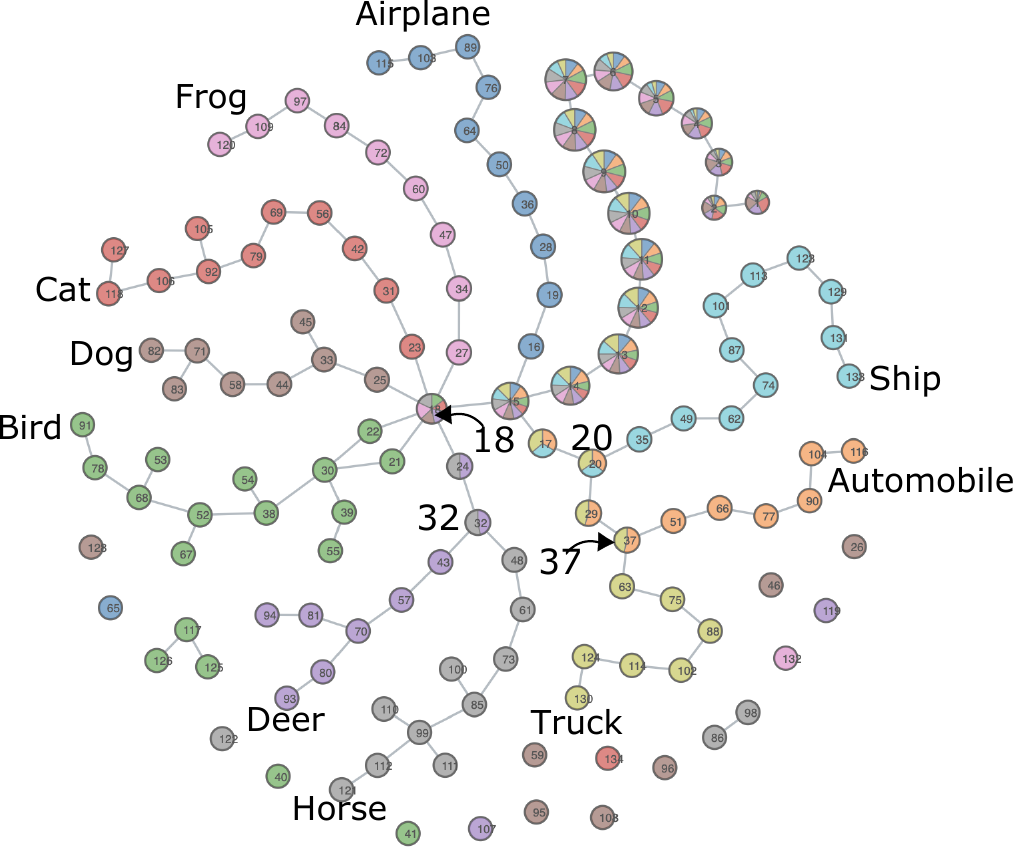}
\vspace{-3mm}
\caption{The mapper graph of the activation vectors. For DBSCAN, $\epsilon = 8.71$, $minPts = 5$. For mapper graphs, $ n= 40$, $p =0.2$.}
\label{fig:cifar10-new}
\vspace{-3mm}
\end{figure}

We compute the pairwise distance matrix using all $512$ dimensions, and use $L_2$-norm as the filter function. 
As shown in~\autoref{fig:cifar10-new}, the resulting mapper graph highlights the separation of image classes at the last layer of a trained neural network (ResNet-18) with high classification accuracy. 
By drawing a pie chart on top of each node, the proportions of categories within each node are clear. 
The size of each node reflects the number of points within the node (cluster).

The resulting mapper graph not only clusters images from each class into a separate branch, but also highlights the relationship among the different classes. 
For example, {\tool} highlights the observation from~\cite{RathoreChalapathiPalande2021}. 
A branch of nodes containing the deer and horse images first emerged from the branching node 18, which contains images from several classes. Then, the two classes are separated into two branches at node 32. The branching order indicates that the deer and horse images are more similar than images from other class categories. 
Similarly for the automobile and truck images, a branch containing images from both categories first emerged from branching node 20, and then the two categories were separated from each other at branching node 37, indicating the automobile and truck images are more similar than other images.

\begin{figure}[!ht]
\centering
\includegraphics[width=0.99\columnwidth]{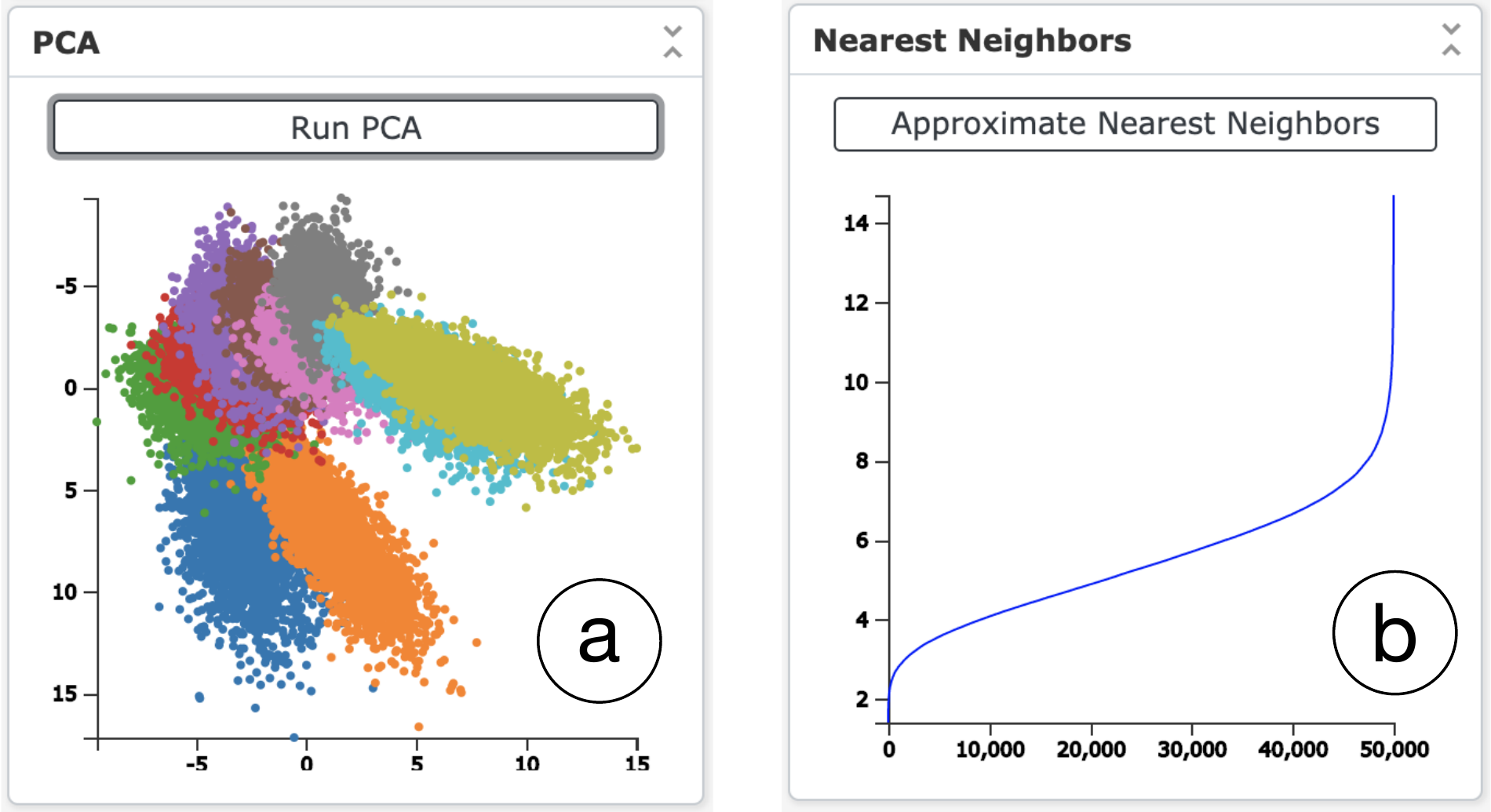}
\vspace{-2mm}
\caption{Adding two additional analysis modules. (a) The result of a PCA module. (b) Computing the distance distribution of the fifth nearest neighbors for all points.}
\label{fig:cifar10-modules}
\vspace{-2mm}
\end{figure}

Using {\tool}, we can easily perform additional analysis to further advance our understanding of the dataset. 
We can extend the tool by adding a PCA analysis module and a module for visualizing the distribution of nearest neighbor distances. 
\autoref{fig:cifar10-modules}a applies PCA to the activation vectors. 
The colors correspond to the 10 class categories used in the mapper graph. 
Compared to the mapper graph, the PCA projection of the activation vectors does not separate the classes well, and the relationship between different classes is not well depicted. 

In addition, for such a large dataset, parameter tuning can be difficult and time consuming. We demonstrate how to add a new module for tuning the $\epsilon$ parameter in DBSCAN by creating a module in the expert mode. 
We use the Python library \emph{PyNNDescent} to approximate a nearest neighbor search for the point cloud data, sort the distances of the $k$-th nearest neighbor, and plot the distance distribution to figure out the most appropriate $\epsilon$ value. 

Our objective is to have, with the right $\epsilon$ value, a maximum number of points with at least  $k$ neighbors, with the maximum distance not too large to include too many neighbors. 
For the CIFAR-10 dataset, we chose $k=5$, and the distance distribution of the fifth nearest neighbors is shown in~\autoref{fig:cifar10-modules}b. The line plot shows that an $\epsilon$ value around 8 is most appropriate.
In DBSCAN clustering, the parameter $\epsilon$ is the maximum distance between two points for them to be considered as neighbors. 
If $\epsilon$ is too small, the number of neighbors for most points will be less than the minimum number of points to be clustered together, and most points will be considered as noise, resulting in a suboptimal  clustering. 
If $\epsilon$ is too large, then all the points will be considered as neighbors of each other, and thus be assigned to the same cluster, which will also be suboptimal. 
The nearest neighbor module in \autoref{fig:cifar10-modules}b, therefore, helps the users choose the optimal parameter for DBSCAN.

\subsection{Exploring Breast Cancer Data}
Our third use case is to provide alternative ways to explore the results from a breast cancer study~\cite{LumSinghLehman2013}. 
Lum \etal~\cite{LumSinghLehman2013} utilized the mapper algorithm to identify subgroups in breast cancer patients. 
Using {\tool}, the key point is that we can consider alternative configurations of the mapper algorithm to explore these breast cancer datasets and obtain similar insights. 
Due to the extendability of the tool, we can further provide in-depth regression analysis to identify possible factors that are highly correlated with patient survival. 

\begin{figure}[!ht]
\centering
\includegraphics[width=0.99\columnwidth]{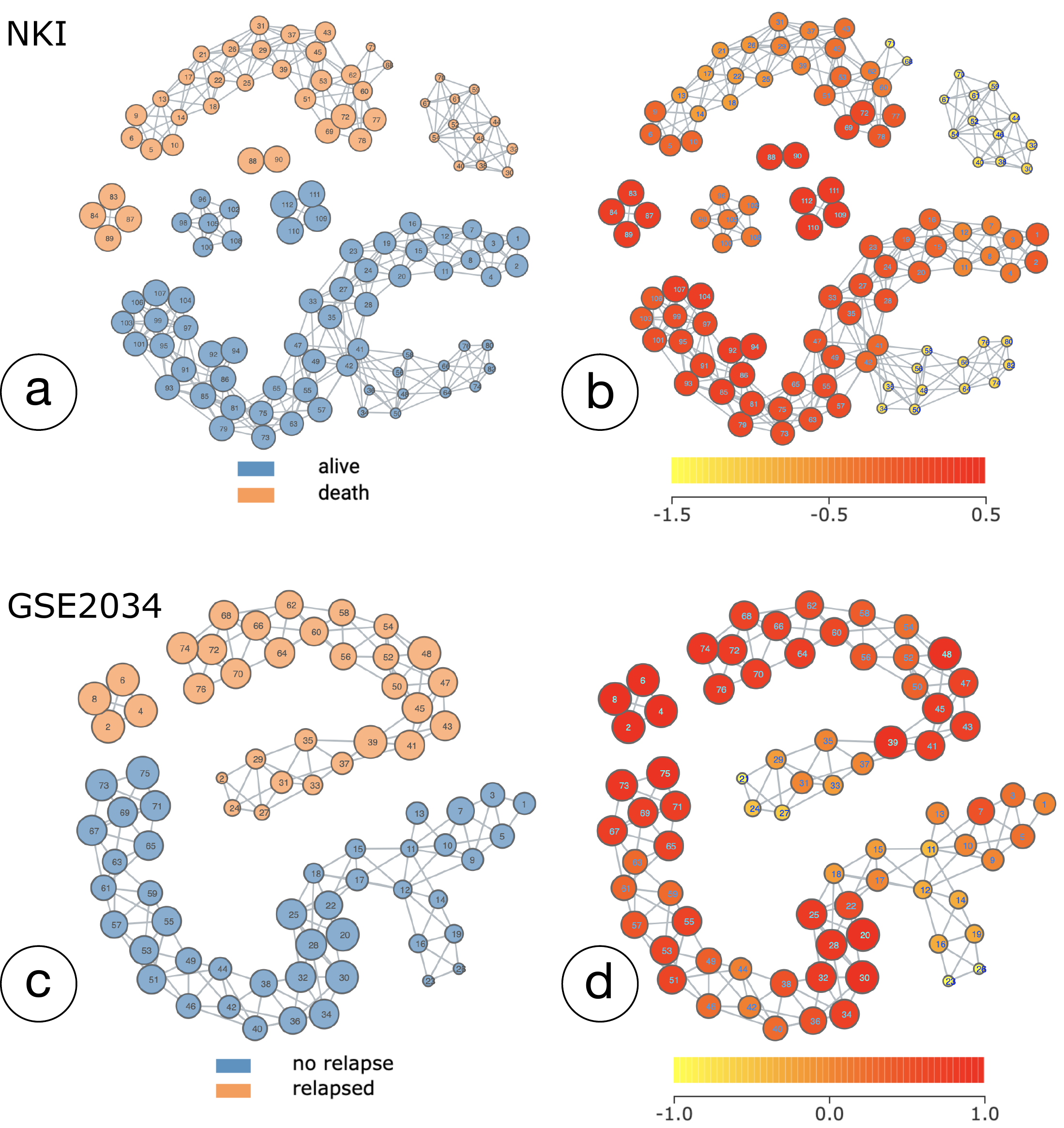}
\vspace{-2mm}
\caption{(a-b) NKI mapper graphs. For DBSCAN, $\epsilon = 15$, $minPts = 2$. For mapper graph: $f_1=L_\infty$-norm, $n_1=78$, $p_1=65\%$; $f_2=event\_death$, $n_2=10$, $p_2=68\%$. (c-d)  GSE2034 mapper graphs. For DBSCAN, $\epsilon = 0.45$, $minPts = 2$. For mapper graph: $f_1=L_\infty$-norm, $n_1=37$, $p_1=72\%$; $f_2=relapse$, $n_2=5$, $p_2=50\%$. Each dimension is normalized by a min-max scale.}
\label{fig:breast_cancer_mapper}
\vspace{-4mm}
\end{figure}

\para{Insight discovery.}
We first discuss insight discovery with alternative configurations of the mapper algorithm.
We use the two breast cancer datasets studied by Lum \etal~\cite{LumSinghLehman2013}, referred to as the \emph{NKI} dataset~\cite{VeerDaiVijver2002}, and the \emph{GSE2034}  dataset~\cite{WangKlijnZhang2005}.

The \emph{NKI} dataset contains information from 272 breast cancer patients (rows). 
For each patient, two types of variables (columns) are recorded: the first type of variables contains 1554 gene expression levels, and the second type of variables consists of other medical records or physiological measures. 
The physiological measure columns include $event\_death$ (whether a patient survived or not), $survival\_time$, $recurrence\_time$, $chemo$ (whether a patient received chemotherapy), $hormonal$ (whether a patient received hormonal therapy), $amputation$ (whether forequarter amputation has been used), $hist\_type$ (histological type), $diam$ (diameter of the tumor), $posnodes$ (number of nodes), $grade$ (cancel level), $angio\_inv$ (to what degree the cancer invaded blood vessels and lymph vessels), and $lymph\_infil$ (level of lymphocytic infiltration).
 
To compute the mapper graph of the \emph{NKI} dataset, we inherit some parameter configurations from~\cite{LumSinghLehman2013}, with the exception that we use DBSCAN as our clustering algorithm instead of the single-linkage clustering employed by Lum \etal~in ~\cite{LumSinghLehman2013}. 
Subsequently, we use slightly different parameters for the number of intervals $n$ and the amount of overlap $p$ that is adaptive to DBSCAN.  
We take the $1500$ mostly varying genes to form a point cloud in $1500$ dimensions, compute their Euclidean pairwise distance matrix, and construct a 2D mapper graph using $L_\infty$-norm and the response variable $event\_death$ as its filter functions. 

The \emph{GSE2034} dataset, on the other hand, consists of gene expression levels of $22283$ genes from $286$ patients. 
Instead of recording the survival data, this dataset provides a variable $relapse$ to indicate whether the patient suffered a relapse. 
We take the top $10$ most varying genes to compute the pairwise distance matrix, and construct a 2D mapper graph using $L_\infty$-norm and the $relapse$ variable as its filter functions.

Lim \etal~\cite{LumSinghLehman2013} used mapper graphs to study  subgroups of breast cancer patients. In most cases, the expression level of the estrogen receptor gene (ESR1) is positively correlated with the prognosis. Patients with high ESR1 levels usually have a better prognosis and are more likely to survive than patients with low ESR1 levels. However, among all the patients with high ESR1 levels are subgroups having poor clinical outcomes. Patients who had low ESR1 levels but survived were also identified over the years. Researchers have studied such subgroups using certain experimental data~\cite{TeschendorffMiremadiPinder2007, SorliePerouTibshirani2001, PerouSorlieEisen2000}. However, the challenge is to identify subgroups under more general settings, such as data from different sets of patients that are collected at different times.

We therefore apply {\tool} to both \emph{NKI} and \emph{GSE2034} datasets using slightly different parameter configurations in comparison to~\cite{LumSinghLehman2013}. 
The resulting mapper graphs are shown in~\autoref{fig:breast_cancer_mapper}. 
It is interesting to observe that the two resulting mapper graphs consist of similar structure for the survivor/non-relapse patients; and they share similar (but not identical) structures in comparison to the results from~\cite{LumSinghLehman2013}. 
In each graph, the blue connected component on the bottom contains a branch of nodes with low average ESR1 expression levels, thus defining a subgroup of survivor/nonrelapse patients. 
The result shows that we are able to visually identify similar subgroup structures under two datasets generated from totally different experimental settings. 

\para{In-depth exploration.}~Furthermore, we can easily extend {\tool} to perform in-depth analysis of the breast cancer datasets (not discussed in~\cite{LumSinghLehman2013}).   

\begin{figure}[!ht]
\centering
\includegraphics[width=0.99\columnwidth]{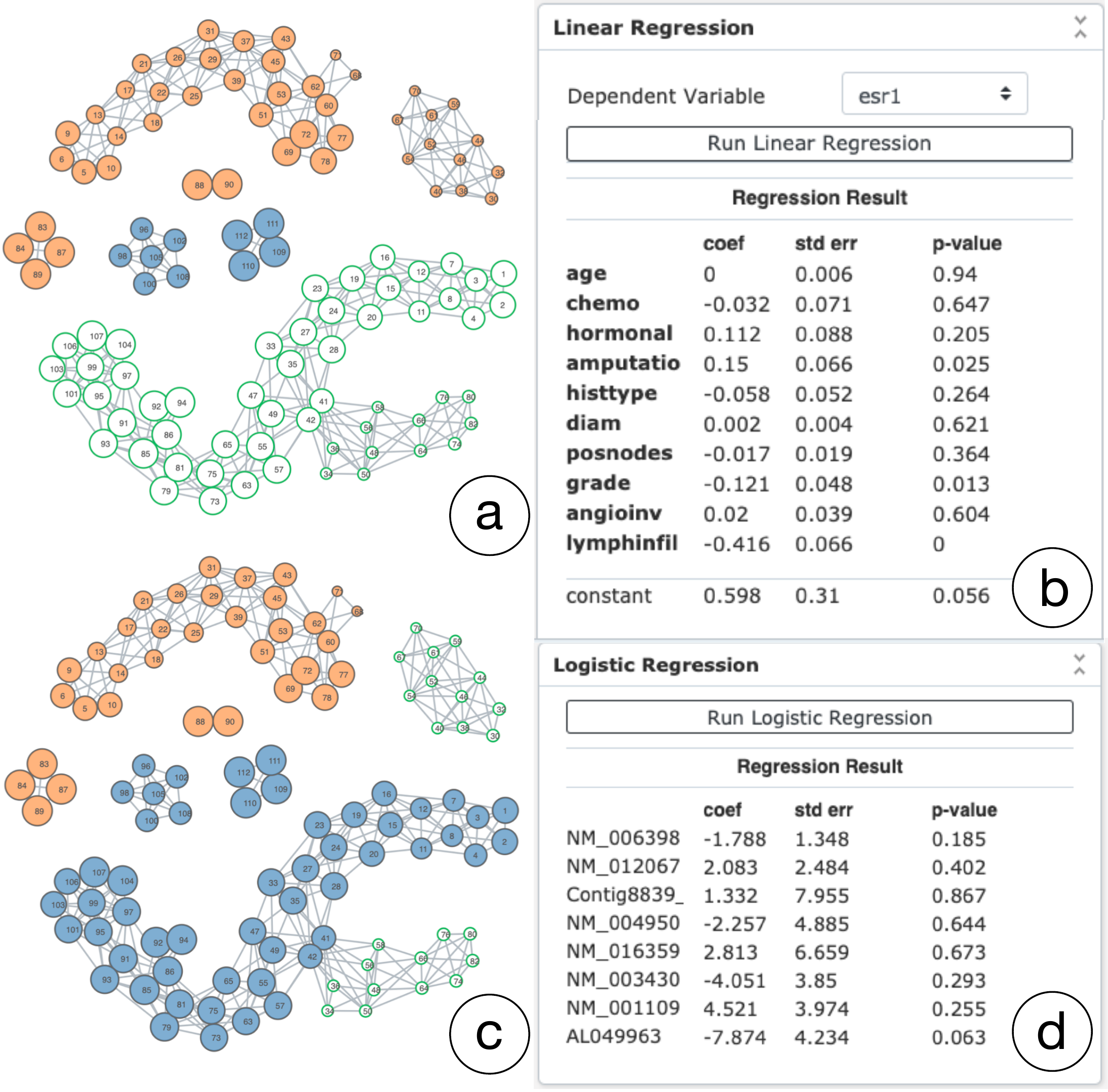}
\vspace{-1mm}
\caption{In-depth exploration of the \emph{NKI} dataset. (a) White nodes with green edges are selected to perform linear regression on ESR1 levels. (b) Linear regression result. (c) White nodes with green edges are selected to perform logistic regression on $event\_death$. (d) Logistic regression result.}
\label{fig:nki_regression}
\vspace{-2mm}
\end{figure}

Since the \emph{NKI} dataset contains medical records and physiological measures information about the patients, we can make use of the analysis modules to explore interesting subsets of the mapper graph. 
As illustrated in~\autoref{fig:nki_regression}a, we first consider the clusters of the largest connected component among the survivors (the green selected clusters). 
Since there is a subgroup of low ESR1 patients in these clusters, we can easily apply the existing linear regression module to these clusters. 
Specifically, we are interested in understanding what variables have statistically significant effects on the expression levels of ESR1 without affecting the patient survival. 

The result is shown in~\autoref{fig:nki_regression}b. 
Under the significant level ($p$-value) of $0.05$, the variables $amputation$, $grade$, and $lymph\_infil$ are significantly correlated with the expression levels of ESR1. 
Recall that the $amputation$ variable indicates whether the patient has received the forequarter amputation treatment, the $grade$ variable indicates the stage of the cancer, and the $lymph\_infil$ variable indicates the level of lymphocytic infiltration. 

We explore which genes affect the survival of patients with low levels of ESR1. Since $event\_death$ is a binary variable, we add a new module to perform logistic regression. 
We include the top $10$ genes that are selected using a recursive feature elimination. 
The result is shown in~\autoref{fig:nki_regression}d.  
Further analysis based on these regression results would be an interesting avenue to explore in a follow-up study. 

A subset of results for these use cases cannot be easily reproduced using other frameworks. {\tool} supports the application of ML techniques to a subset of nodes interactively. 
Therefore, in the breast cancer example, we can easily apply regression to the identified subgroups using path-based and/or  component-based node selection. 
\emph{giotto-tda} or \emph{KeplerMapper} will require more coding effort to achieve the same result. 
Neither \emph{giotto-tda} nor \emph{KeplerMapper} support pie charts on top of the nodes for discrete variables; therefore, it is hard to identify the branching points in both COVID-19 and the neuron activation examples. 
Overall, {\tool} provides more interactivity and flexibility, and is less time intensive for nonspecialists exploring high-dimensional data with TDA.

\section{Conclusion and Discussion}
\label{sec:discussion}

In this paper, we present {\tool}, an interactive, extendable, and scalable toolbox for the visual exploration of high-dimensional data using the mapper graph. 
It supports computation and interactive exploration of mapper graphs. 
It is easily extendable, where both novice and expert users can add machine learning and visualization modules with a few changes to a json file or with a few lines of codes.  
\myedit{Since it provides low-code development environment, it also enables quickly checking if TDA is a viable tool for a given application.}
Its command line API can compute a mapper graph of 1 million points in 256 dimension in less than 3 minutes, which is roughly 4 times faster than the state-of-the-art single processor vanilla implementation. 
Its GPU implementation of the mapper graph computation provides an additional 2-fold acceleration in comparison to its CPU counterpart. 
\myedit{We have shown in~\autoref{table:compare} that {\tool} provides a large number of unique features compared with the state-of-art. We have demonstrated the usefulness of such features via three use cases.} 
We discuss a few possible extensions of {\tool} together with challenges and opportunities.  

\para{Pushing the scalability boundary.}
The scalability boundary of mapper graph computation can be pushed even further, especially for larger datasets with more than 10 million points.  
Assuming sufficient storage and memory, one of the limiting factors is the clustering step during the mapper construction.
In this paper, we are able to speed up the clustering process by parallelizing the distance computation on a single (multicore) CPU, as well as a single GPU. 
One obvious avenue is to distribute the distance computation across multiple GPUs.
This task is nontrivial since extensive testing is needed to balance the trade-off between moving data from the CPU to multiple GPUs and merging the results across GPUs.

GPU memory will also become a bottleneck. As the data size increases, the distance matrix grows quadratically.
The support for large amounts of memory, on the order of hundreds of gigabytes, is limited on GPUs.
We also notice in our implementation, with a large number of intervals, the overhead from moving matrices from CPU to GPU and back increases since this operation is done per interval.
When the amount of input data is small, initializing the necessary CUDA kernels is also a severe overhead. 
As more GPUs are added, this overhead will become more pronounced.

Although DBSCAN is the primary clustering algorithm used in {\tool}, an algorithm built for high-performance computing, such as DBSCAN++~\cite{JangJiang2019}, could provide additional runtime benefits.
Another promising area is using an approximate nearest neighbor library such as \emph{PyNNDescent}~\cite{DongMosesLi2011} instead of computing a the full distance matrix. 
\myedit{Finally, datasets of different sizes and dimensionality warrant employing different sets of optimizations. Future work on the tool would entail automatic inference of the optimization strategies based on the input data and system configuration.}

\para{Improving extendability for novice and expert users.}
One of the strengths of {\tool} is that it allows users to extend the current analysis and visualization capabilities by adding modules that interface with \emph{scikit-learn}, which also leaves plenty of room for improvement. 
Many common data analysis libraries follow the well-established API guidelines set forth by \emph{scikit-learn}.
Because of this standardization, implementing new libraries is  straightforward pragmatically, especially under the expert user setting. 
Making such extensions accessible for novice users through configuration files is nontrivial and is left for future work. 

\para{Parameter selection for the mapper algorithm.}
\myedit{Parameter selection for mapper algorithm is a challenging open problem. 
We use a ``best practice'' for parameter selection for the mapper algorithm commonly employed by the practitioners. That is, finding a range of parameters where the graph produces stable structures which is enabled by the ability to change parameters on-the-fly in {\tool} and immediately visualize and explore the resulting structures. Carri{\'e}re \etal~\cite{CarriereMichelOudot2018} provided theoretical results for automatic parameter tuning. However, their theoretical results require the data to adhere to certain statistical assumptions that are often not applicable to real world datasets. Automatic parameter tuning in practice  for the mapper algorithm remains an open problem to be tackled by the TDA community.}

\vspace{-1mm}
\acknowledgments{
The authors wish to thank Nathaniel Saul for comments on the initial prototype of our tool; Anantharaman Kalyanaraman, Methun Kamruzzaman, Bala Krishnamoorthy, and Umberto Lupo for valuable 
suggestions and discussions. 
This work is partially supported by NSF IIS-1513616 and NSF DBI-1661375.}


\clearpage
\newpage
\appendix
\onecolumn
\section*{Supplement: Mapper Interactive: A Scalable, Extendable, and Interactive Toolbox for the Visual Exploration of High-Dimensional Data}
\begin{table*}[!ht]
	\centering
	\begin{tabular}{|l|r|r|r|r|r|r|r|r|}
		\hline
		\textbf{Dataset}  &   \textbf{MI} &   \textbf{GT} &   \textbf{KM} &  \textbf{GT*} &  \textbf{KM*} & \textbf{MI/KM} & \textbf{GT*/GT} & \textbf{KM*/KM}  \\ \hline
		\emph{ImageNet} & 3.64 & 1.82 & 2.35 & 6.67 & 2.62  & $1.55\times$ & $3.66\times$ & $1.11\times$ \\ \hline
		\emph{Cifar}    & 1.23 & 0.57 & 0.74 & 1.90 & 1.16 & $1.66\times$ & $3.33\times$ & $1.57 \times$ \\ \hline
		\emph{Random}   & 0.51 & 0.34 & 0.37 & 1.35 & 0.71 & $1.38 \times$ & $3.97\times$ & $1.92 \times$ \\ \hline
	\end{tabular}	
	\vspace{-2mm}
	\caption{Peak memory usage (in Gigabytes) on three datasets, each
with 100K points.}
\label{table:memory-comparison}
\end{table*}

\begin{table*}[!ht]
	\centering
	\begin{tabular}{|l|m{0.04\textwidth}|m{0.05\textwidth}|m{0.05\textwidth}|m{0.05\textwidth}|m{0.05\textwidth}|m{0.05\textwidth}|m{0.07\textwidth}|m{0.07\textwidth}|m{0.07\textwidth}|}
	\hline
	\textbf{\# Pts}  & \textbf{Int} & \textbf{MI} & \textbf{GT} & \textbf{KM} & \textbf{GT*} & \textbf{KM*} & \textbf{KM/MI} & \textbf{GT/GT*} & \textbf{KM/KM*}	\\ \hline
	\rowcolor{aliceblue}\multicolumn{10}{|c|}{\emph{ImageNet} dataset} \\ \hline
	$1 \times 10^2$	&	5	&	0.01	&	0.27	&	0.21	&	0.1	&	0.21	&	21.0 $\times$ & 2.7 $\times$ & 1.0 $\times$	 \\ \hline
	$1 \times 10^3$	&	10	&	0.05	&	0.79	&	0.65	&	0.11	&	0.65	&	13.0 $\times$ & 7.18 $\times$ & 1.0 $\times$	 \\ \hline
	$1 \times 10^4$	&	20	&	0.99	&	4.83	&	5.52	&	0.69	&	2.42	&	5.58 $\times$ & 7.0 $\times$ & 2.28 $\times$	 \\ \hline
	$1 \times 10^5$	&	100	&	22.08	&	135.29	&	143.04	&	13.82	&	22.26	&	6.48 $\times$ & 9.79 $\times$ & 6.43 $\times$	 \\ \hline
	$3 \times 10^5$	&	200	&	99.74	&	562.19	&	578.84	&	60.99	&	88.03	&	5.8 $\times$ & 9.22 $\times$ & 6.58 $\times$	 \\ \hline
	
	\rowcolor{aliceblue}\multicolumn{10}{|c|}{\emph{Cifar} dataset}    \\ \hline
	$1 \times 10^2$	&	5	&	0.01	&	0.18	&	0.31	&	0.1	&	0.31	&	31.0 $\times$ & 1.8 $\times$ & 1.0 $\times$	 \\ \hline
	$1 \times 10^3$	&	10	&	0.05	&	0.45	&	0.94	&	0.11	&	0.96	&	18.8 $\times$ & 4.09 $\times$ & 0.98 $\times$	 \\ \hline
	$1 \times 10^4$	&	20	&	0.82	&	3.0	&	3.78	&	0.46	&	2.07	&	4.61 $\times$ & 6.52 $\times$ & 1.83 $\times$	 \\ \hline
	$1 \times 10^5$	&	100	&	12.73	&	43.9	&	57.51	&	7.45	&	15.71	&	4.52 $\times$ & 5.89 $\times$ & 3.66 $\times$	 \\ \hline
	$1 \times 10^6$	&	500	&	265.86	&	932.77	&	1182.85	&	171.97	&	214.24	&	4.45 $\times$ & 5.42 $\times$ & 5.52 $\times$	 \\ \hline
	$3 \times 10^6$	&	1500	&	931.73	&	3392.51	&	3740.08	&	1025.79	&	802.77	&	4.01 $\times$ & 3.31 $\times$ & 4.66 $\times$	 \\ \hline
	\rowcolor{aliceblue}\multicolumn{10}{|c|}{\emph{Random} dataset}   \\ \hline
	$1 \times 10^2$	&	5	&	0.01	&	0.1	&	0.51	&	0.1	&	0.52	&	51.0 $\times$ & 1.0 $\times$ & 0.98 $\times$	 \\ \hline
	$1 \times 10^3$	&	10	&	0.02	&	0.43	&	0.73	&	0.12	&	0.74	&	36.5 $\times$ & 3.58 $\times$ & 0.99 $\times$	 \\ \hline
	$1 \times 10^4$	&	20	&	0.62	&	1.64	&	2.47	&	0.5	&	2.15	&	3.98 $\times$ & 3.28 $\times$ & 1.15 $\times$	 \\ \hline
	$1 \times 10^5$	&	100	&	9.42	&	22.17	&	32.6	&	5.56	&	13.25	&	3.46 $\times$ & 3.99 $\times$ & 2.46 $\times$	 \\ \hline
	$1 \times 10^6$	&	500	&	185.2	&	289.11	&	389.7	&	113.16	&	154.86	&	2.10 $\times$ & 2.55 $\times$ & 2.52 $\times$	 \\ \hline
	$1 \times 10^7$	&	10000	&	1738.57	&	OOM	&	4556.77	&	OOM	&	1963.23	&	2.62 $\times$ & OOM  & 2.32 $\times$	 \\ \hline
\end{tabular}
\vspace{-2mm}	
\caption{Runtime comparison (in seconds) of our implementation vs \emph{KeplerMapper} (KM) and \emph{giotto-tda} (GT) on the \emph{ImageNet}, \emph{Cifar}, and \emph{Random} datasets, respectively. \textbf{Int}: intervals. \textbf{OOM}: out of memory. \textbf{N/A}: not available. \textbf{KM/MI}, \textbf{GT/GT*}, \textbf{KM/KM*}: speed up factors.}
\label{table:runtime-all}
\end{table*}

\begin{table*}[!ht]
    \centering    
    \begin{tabular}{|l|r|r|r|r|r|}
    \hline
        \textbf{Data Size} &  \textbf{Intervals} & \textbf{CPU Version} &        \textbf{GPU Version} & \textbf{CPU/GPU}\\ \hline
        \rowcolor{aliceblue}\multicolumn{5}{|c|}{\emph{ImageNet} dataset} \\ \hline
        $1 \times 10^2$ &       5 &   0.12 &   0.05 & 2.40$\times$ \\ \hline
        $1 \times 10^3$ &      10 &   0.67 &   0.06 & 11.17$\times$\\ \hline
        $1 \times 10^4$ &      20 &   2.54 &   0.80 & 3.18$\times$\\ \hline
        $1 \times 10^5$ &     100 &  22.32 &  12.93 & 1.73$\times$\\ \hline
        $3 \times 10^5$ &     500 &  69.88 &  36.22 & 1.93$\times$\\ \hline
         \rowcolor{aliceblue}\multicolumn{5}{|c|}{\emph{Cifar} dataset} \\ \hline
        $1 \times 10^2$ &       5 &     0.042 &    3.95 & 0.01$\times$ \\ \hline
        $1 \times 10^3$ &      10 &     0.59 &    0.08 & 7.38$\times$\\ \hline
        $1 \times 10^4$ &      20 &     2.17 &    0.53 & 4.09$\times$\\ \hline
        $1 \times 10^5$ &     100 &    20.20 &    9.18 & 2.20$\times$ \\ \hline
        $1 \times 10^6$ &     500 &   331.58 &  202.87 & 1.63$\times$ \\ \hline
        $3 \times 10^6$ &    1500 &  1142.82 &  753.37 & 1.52$\times$\\ \hline
       \rowcolor{aliceblue}\multicolumn{5}{|c|}{\emph{Random} dataset (128-dimension)} \\ \hline
        $1 \times 10^2$ &       5 &     0.05 &    3.97 & 0.01$\times$\\ \hline
        $1 \times 10^3$ &      10 &     0.71 &    0.03 & 23.67$\times$\\ \hline
        $1 \times 10^4$ &      20 &     2.21 &    0.36 & 6.14$\times$\\ \hline
        $1 \times 10^5$ &     100 &    15.07 &    5.09 & 2.96$\times$\\ \hline
        $1 \times 10^6$ &     500 &   256.36 &   122.64 & 2.09$\times$ \\ \hline
        $1 \times 10^7$ &    1500 &  5234.30 &  4269.83 & 1.23$\times$\\ \hline
    \end{tabular}
\vspace{-2mm}    
\caption{Runtime comparison (in seconds) of our implementation on CPU vs GPU using three testing datasets. \textbf{CPU/GPU}: speed up factors.}
    \label{table:GPU-all}
\end{table*}

\end{document}